\documentclass[a4paper,12pt]{JHEP3}

\usepackage{amsmath}
\usepackage{amssymb}
\usepackage{latexsym}
\usepackage[dvips]{graphicx}
\usepackage{bbold}
\usepackage{epsfig}
\usepackage{amsfonts}

\newcommand{\beqa}{\begin{eqnarray}}
\newcommand{\eeqa}{\end{eqnarray}}
\newcommand{\bea}{\begin{eqnarray}}
\newcommand{\eea}{\end{eqnarray}}

\newcommand{\be}{\begin{equation}}
\newcommand{\ee}{\end{equation}}
\newcommand{\half}{\frac{1}{2}}

\newcommand{\mcR}{{\mathcal R}}

\newcommand{\mcK}{{\mathcal K}}

\numberwithin{equation}{section}

\title{On a class of 4D K\"ahler bases \\ and AdS$_5$ supersymmetric Black Holes}
\author{Pau Figueras\\Departament de F\'\i sica Fonamental, and, \\ C.E.R en Astrof\'\i sica, F\'\i sica de Part\'\i cules i Cosmologia,\\ Universitat de Barcelona, Diagonal 647, E-08028 Barcelona, Spain \\ E-mail: \email{pfigueras@ffn.ub.es}}
\author{Carlos A R Herdeiro and Filipe Paccetti Correia\\ Centro de F\'\i sica do Porto,\\ Faculdade de Ci\^encias da Universidade do Porto,\\ Rua do Campo Alegre, 687, 4169-007 Porto, Portugal \\ E-mails: \email{crherdei@fc.up.pt}; \email{paccetti@fc.up.pt}}

\abstract{We construct a class of toric K\"ahler manifolds, $\mathcal{M}_4$, of real dimension four, a subset of which corresponds to the K\"ahler bases of all known 5D asymptotically $AdS_5$ supersymmetric black-holes. In a certain limit, these K\"ahler spaces take the form of cones over Sasaki spaces, which, in turn, are fibrations over toric manifolds of real dimension two. The metric on $\mathcal{M}_4$ is completely determined by a single function $H(x)$, which is the conformal factor of the two dimensional space. We study the solutions of minimal five dimensional gauged supergravity having this class of K\"ahler spaces as base and show that in order to generate a five dimensional solution $H(x)$ must obey a simple sixth order differential equation. We discuss the solutions in detail, which include all known asymptotically $AdS_5$ black holes as well as other spacetimes with non-compact horizons. Moreover we find an infinite number of supersymmetric deformations of these spacetimes with less spatial isometries than the base space. These deformations vanish at the horizon, but become relevant asymptotically.}

\preprint{hep-th/0608201}


\begin{document}

\section{Introduction}
The study of supersymmetric, asymptotically $AdS_5$, black hole type solutions has been a topic of interest over the last few years. The reason is, at least, two-folded. 

On the one hand, the $AdS/CFT$ correspondence \cite{Maldacena:1997re,Witten:1998qj,Gubser:1998bc} is best understood in the case of $AdS_5/CFT_4$, the latter corresponding to $\mathcal{N}=4$, $D=4$, Super Yang Mills Theory. Thus, the discovery of explicit black hole solutions on the gravity side, especially supersymmetric ones \cite{Gutowski:2004ez,Gutowski:2004yv,Chong:2005da,Chong:2005hr,Kunduri:2006ek,Chong:2006zx}, opened the possibility of a microscopic description of their entropy using the correspondence. This programme, which has been pursued by several authors \cite{Kinney:2005ej,Berkooz:2006wc,Silva:2006xv,Kim:2006he}, has not been fully successful yet. One of the puzzling points is that there is a discrepancy in the counting of parameters of the most general supersymmetric black hole solution known and the number of parameters expected from the CFT side. The most general supersymmetric black holes known are characterised by 5 parameters, two angular momenta and three charges, which are not all independent. In particular this leads to an ambiguity in writing an entropy formula, since one can take different choices for the 4 independent parameters (see \cite{Kim:2006he} for a promising proposal for such formula). From the CFT viewpoint one would, in principle, expect these 5 parameters to be independent. Thus it would be interesting to further explore more general black hole solutions, which were also conjectured to exist in \cite{Berkooz:2006wc}.

On the other hand, five dimensional gravity has provided the first examples of black hole non-uniqueness. The most striking example is the existence of black holes with different horizon topologies and with the same conserved charges \cite{Emparan:2001wn}. This discrete non-uniqueness was first found for non-supersymmetric, asymptotically flat configurations, and was generalised to supersymmetric solutions of non-minimal five dimensional supergravity in \cite{Elvang:2004ds,Bena:2004de,Gauntlett:2004qy}. In the minimal supergravity case, however, these solutions reduce to a three parameter family, where uniqueness holds. The physical parameters are charge, and the two independent angular momenta of $SO(4)$, $J_1$ and $J_2$. Interestingly, there is no region of physical parameters where black rings and black holes with a (topologically) spherical horizon may coexist \cite{Elvang:2004rt}, which legitimates the string theory counting of black hole micro-states for the BMPV black hole performed in \cite{Breckenridge:1996is}. In fact, a topologically spherical horizon requires $J_1=J_2$, whereas supersymmetric black rings require $J_1\neq J_2$. By contrast, in the asymptotically $AdS$ case, there are already supersymmetric black holes with a (topologically) spherical horizon and two independent angular momenta; moreover, no supersymmetric black ring is yet known. The spectrum of supersymmetric black holes is, therefore, quite different in the $AdS$ case, and some black objects, like black rings, are yet to be found. Again we conclude it would be interesting to further explore the spectrum of asymptotically $AdS_5$ black objects.

Another suggestion of a richer structure for asymptotically $AdS$ black holes is the inability to derive the general form for the geometry and topology of the horizon \cite{Gutowski:2004yv} using the approach originally taken in \cite{Reall:2002bh} for the asymptotically flat space case. For the $AdS$ case this method used the construction of \cite{Gauntlett:2003fk} for general supersymmetric solutions of five dimensional gauged supergravity, together with the assumption of the existence of physical horizon, which must be a Killing horizon. Whereas in the asymptotically flat case, the resulting equations could be integrated to obtain two theorems concerning the geometry of the solution near the horizon and the horizon itself \cite{Reall:2002bh}, in the asymptotically $AdS_5$ case, no general answer could be obtained  and only some special solutions were found \cite{Gutowski:2004yv}.

In this paper, therefore, we will pursue a different programme. Our starting point will be the general equations for the timelike supersymmetric solutions of $D=5$ minimal gauged  SUGRA found by Gauntlett and Gutowski \cite{Gauntlett:2003fk}. Their construction starts off with a choice of a four dimensional K\"ahler manifold - the \textit{base} space -; once chosen the base, the full solution follows from a series of constraint equations that involve the base. We will make some assumptions for the base in order for the full solution to be compatible with having a finite size horizon. These assumptions lead to the conclusion that the base should behave, near the putative horizon, as a K\"ahler cone. In the most general framework, this K\"ahler cone admits a single $U(1)$ action. However we will focus on a special case in which the base admits a $U(1)^2$ action, which is more tractable. Any K\"ahler cone is a cone over a  three dimensional Sasaki space which is always a fibration over a two dimensional space. For the case under study herein, the four dimensional base is described by a single function which is the conformal factor of the two dimensional space. The full base (i.e. not only close to the putative horizon) is obtained by a plausible ansatz for the K\"ahler potential, and it turns out to be a simple generalisation of the K\"ahler base studied earlier in \cite{Kunduri:2006ek}.

Our assumptions are not fully general, and, in particular, we expect a possible asymptotically $AdS_5$ black ring not to obey them, as explained in section \ref{blackring}. Still, our assumptions are sufficiently general to give i) all known black hole solutions in a simple way and somewhat suggestive that these are all the possible asymptotically $AdS_5$ solutions with a (topologically) spherical horizon and a $U(1)^2$ spatial isometry group; ii) some other previously known and unknown solutions with non-compact or singular horizons; iii) an infinite set of new solutions which correspond to deformations of the known black hole solutions (or even $AdS_5$). Such deformations vanish at the horizon but become relevant asymptotically. These latter solutions have some hitherto unexplored features: the five dimensional spacetime has less spatial isometries than the base and, if one writes $AdS_5$ in coordinates that reduce to static coordinates in the absence of deformations, these solutions are time dependent.  

It was argued in \cite{Gauntlett:2003fk} that any K\"ahler base should originate a five dimensional solution of $D=5$ minimal gauged SUGRA. However, one other result of our analysis is that some K\"ahler bases do not originate any five dimensional solution. Also, we give explicit examples of K\"ahler bases with a finite number of parameters (that could be even zero) which originate an infinite parameter family of five dimensional solutions, a fact already anticipated, albeit not explored, in \cite{Gauntlett:2003fk}.

This paper is organised as follows. In section 2 we review the formalism in \cite{Gauntlett:2003fk} for supersymmetric solutions of minimal gauged supergravity in $D=5$. In section 3 we make some remarks about co-homogeneity two K\"ahler manifolds, which will be at the heart of the present paper. In section 4, we note that all known supersymmetric $AdS_5$ black holes are defined by a simple two variable function and use this fact to recover the base space used in their construction. In section 5 we introduce our assumptions for the K\"ahler base, and derive that it behaves as a K\"ahler cone near the putative horizon. We also make a few remarks about the geometrical structure of toric K\"ahler cones and suggest why black rings should violate our assumptions. In section 6 we study the five dimensional solutions in the near horizon limit, and in section 7 we extend the base space away  from the horizon and obtain the equations for the most general five dimensional solutions possessing this base space. The solutions are analysed in section 8. We draw our conclusions in section 9. The appendix makes some comments on non-toric K\"ahler cones, which might be of relevance for finding black holes solutions with less isometries.

\section{Supersymmetric solutions of $D=5$ SUGRA}
The bosonic part of the minimal gauged supergravity theory in five spacetime dimensions is described by the action
\be
{\mathcal{S}}=\frac{1}{16\pi G_5}\int d^5 x\sqrt{-\det{g_{\mu \nu}}}\left(R+12g^2-F^2-\frac{2}{3\sqrt{3}}\tilde{\epsilon}^{\alpha \beta \gamma \mu \nu}F_{\alpha \beta}F_{\gamma \mu} A_{\nu}\right) \ , \label{5dsugra}\ee
where $F=dA$, $\tilde{\epsilon}$ is the Levi-Civita tensor, related to the Levi-Civita tensor density by $\tilde{\epsilon}^{\alpha \beta \gamma \delta \mu}=\epsilon^{\alpha \beta \gamma \delta \mu}/\sqrt{-\det{g_{\mu \nu}}}$, $g$ is the inverse of the $AdS$ radius and we use a `mostly plus' signature. The equations of motion are
\be R_{\mu \nu}+(4g^2)g_{\mu \nu}=2\left(F_{\mu \alpha}F_{\nu}^{\ \alpha}-\frac{1}{6}g_{\mu \nu}F^2\right) \ , \ \ \ \ \ D_{\mu}F^{\mu \nu}=\frac{1}{2\sqrt{3}}\tilde{\epsilon}^{\alpha \beta \gamma \mu \nu}F_{\alpha \beta}F_{\gamma \mu} \ . \label{eqmotn1d5}\ee
In \cite{Gauntlett:2003fk}, it was shown that all supersymmetric solutions of this theory, with a timelike Killing vector field, can be casted in the form
\be
                           ds^2=-f^2(dt+w)^2+f^{-1}ds_B^2(\mcK)\ ,
\label{timelikesolution}
\ee
where the 4d base $ds_B^2(\mcK)$ is K\"ahler and $f$, $w$ do not depend on $t$. Moreover, supersymmetry imposes three additional constraints. Defining the self-dual and anti-self-dual parts of $fdw$ by
\be
                  G^{\pm}=\half f(dw\pm\star_4dw)\ , 
\label{Gpm}
\ee
where $\star_4$ is the Hodge dual on the K\"ahler base, these constraints read
\be
                   f=-\frac{24g^2}{R}\ ,
\label{fricci}
\ee
\be
                  G^+=-\frac{1}{2g}\left[\mcR+\frac{R}{4}J\right]\ ,
\label{Gplus}
\ee
and
\be\label{eq:to_solve1}
                  \nabla^2 f^{-1}=\frac{2}{9}G^{+mn}G^+_{mn}-gf^{-1}G^{-mn}J_{mn}-8g^2f^{-2} \ .
\ee
Given a certain K\"ahler base, we can determine its K\"ahler form $J$ (which is anti-self-dual, i.e. $\star_4 J=-J$, according to the conventions of \cite{Gauntlett:2003fk}),  Ricci form $\mcR$ and Ricci scalar $R$. From these quantities one computes $f^{-1}$, which is proportional to the latter, and $G^+$. Then, from \eqref{eq:to_solve1} we determine $G^-$. In general, $G^-$ may be determined only up to some arbitrary functions from this equation. Combining the most general form allowed for $G^-$ with the form for $G^+$ determined from (\ref{Gplus}), we require, from  (\ref{Gpm}), that
\be
d\left(f^{-1}\left(G^++G^-\right)\right)=0 \ . \label{closed}\ee
The resulting constraints mean that \textit{not all K\"ahler bases may give rise to a non-trivial five dimensional solution}. For the family of K\"ahler bases we study in this paper, this is shown explicitly in section \ref{fullspa}. If $G^-$ can be chosen such that the last equation has a solution, then we can choose an $\omega$ that satisfies (\ref{Gpm}) and find a non-trivial five dimensional solution. The solution is completed by the Maxwell field strength which is given by
\be
F=\frac{\sqrt{3}}{2}d[f(dt+w)]-\frac{G^+}{\sqrt{3}}+\sqrt{3}gf^{-1}J \ . 
\label{eqn:Fstrength}\ee
These solutions preserve at least $1/4$ of the supersymmetry of the full theory, i.e two real supercharges.

\section{Co-homogeneity two K\"ahler bases}\label{seckahler}

Let us consider four (real) dimensions K\"ahler manifolds defined by a K\"ahler potential of the type
\be
                        \mcK=\mcK \left(~|z_1|^2,|z_2|^2~\right) \ ,
\ee
that is, $\mcK$ depends on the complex coordinates $z_i$ only through their squared moduli. The resulting K\"ahler base is \textit{toric} and has co-homogeneity two or lower. There are at least two Killing vector fields on the base space, which are not necessarily Killing vector fields of the five dimensional spacetime too, as we will illustrate in section \ref{secsolutions}. Note, however, that it can be shown that requiring the five dimensional spacetime to admit two $U(1)$ actions implies that the base space should also admit such actions.\footnote{One assumes that these two isometries commute with the supersymmetry Killing vector field. We thank Harvey Reall for this comment.} Thus, the above K\"ahler potential is the most generic one that can lead to axisymmetric black holes (or black rings) with two azimuthal Killing vector fields. As a particular case of a K\"ahler base that can be written in this way let us recall Bergmann space, with the K\"ahler potential 
\be
                        \mcK_{Berg}=-\frac{1}{2g^2}\ln\left(1-|z_1|^2-|z_2|^2\right) \ .
\ee
Bergmann space is the base space of empty AdS$_5$ \cite{Gauntlett:2003fk}.

Let us set our notations and conventions. The metric is obtained from the second derivatives of $\mcK$ ($\mcK_{i{\bar j}}=\partial^2\mcK/\partial{z^i}\partial{{\bar z}^{\bar j}}$), ($i,j=1,2$),
\be
                   ds^2=\mcK_{i{\bar j}}\,dz^i\otimes d{\bar z}^{\bar j}+\mcK_{{\bar i}j}\, d{\bar z}^{\bar i}\otimes dz^j\ ,
\ee
while the K\"ahler 2-form is
\be
                   J=-i\mcK_{i{\bar j}}\,dz^i\wedge d{\bar z}^{\bar j}\ ,
\ee
and the Ricci-form reads
\be
                 \mcR=-i\left(\partial_i\partial_{\bar j}\ln\det(\mcK)\right)\,dz^i\wedge d{\bar z}^{\bar j}\ .
\ee

An useful choice of \emph{real} coordinates is (see, e.g. \cite{Martelli:2005tp})\be
                   z^i=\exp(x^i+i\phi^i)\ ,
\ee 
which leads to a suggestive form of the metric (we denote $G_{ij}\equiv \partial_{x^i}\partial_{x^j}G$ ,  $G_{i}\equiv \partial_{x^i}G$) 
\be
                   ds^2= G_{ij}(x)\left(dx^idx^j+d\phi^id\phi^j\right)\ .
\label{metricGij}
\ee 
Here, the \emph{two}-dimensional metric $G_{ij}$ is the Hessian of $G(x^1,x^2)=\half\mcK(x^1,x^2)$. Due to this relation, we will often refer to $G(x^1,x^2)$ as the ``potential''; $G(x^1,x^2)$ also determines the K\"ahler form
\be
                   J=-\,d\left(G_i(x)\,d\phi^i\right)\ ,
\ee
and the Ricci-form
\be
                   \mcR=-\half d\left[(\partial_i\ln\det(G))\,d\phi^i\right]\ .
\label{ricciformg}\ee
Finally, one can also show that the curvature scalar reads
\be
                   R=-\,G^{ij}\,\partial_i\partial_j\ln\det(G)\ .
\label{riccigij}
\ee

\section{All known $AdS_5$ BH's of minimal gauged SUGRA} 
We will now observe that all co-homogeneity two (or less)  K\"ahler bases, can be described by a family of two-dimensional \emph{curved surfaces} parameterised by $(\exp (x^1) ,\exp (x^2))$. This includes, of course, the bases of all presently known supersymmetric asymptotically anti-de Sitter black holes of five-dimensional minimal gauged supergravity. The most general family was first presented in \cite{Chong:2005hr}, but the base space was first computed in \cite{Kunduri:2006ek}. We believe that these curved surfaces might help building some intuition to search for more general solutions, including black rings.

To be concrete, let us introduce $\rho(x^1,x^2)$, related to the ``potential'' $G$ as
\be\label{eq:def_of_rho}
                      G=-\frac{1}{4g^2}\ln(1-g^2\rho^2) \ .
\ee 
We claim that, to obtain the solutions of \cite{Kunduri:2006ek}, $\rho(x^1,x^2)$ must be determined by
\be\label{eq:BH_curve}
                      \frac{e^{2x^1}}{(g\rho)^{2A_1^2}}+\frac{e^{2x^2}}{(g\rho)^{2A_2^2}}=1\ ,
\ee
where $1\geq A_1^2,A_2^2>0$. For $A_1^2=A_2^2=1$ we recover Bergmann space, for $A_1^2=A_2^2\neq 1$ we retrieve the co-homogeneity one base metric of the Gutowski-Reall black hole \cite{Gutowski:2004ez}, otherwise we have the base space of the $AdS_5$ black holes with two independent angular momenta, first found in \cite{Chong:2005hr} - see the figures: The function $\rho^2(|z_1|^2,|z_2|^2)$ is ploted for the Bergmann manifold, Gutowski-Reall and Chong et al. black holes. $\rho$ varies from $0$ to $1$. For the first two cases $\rho=const$ surfaces are circles, but for the third case  they are ellipses, which, however become a circle as $\rho\rightarrow 1$. This is a manifestation of the asymptotic $AdS_5$ structure of the most general black holes.

\label{bergm}
\DOUBLEFIGURE[t]{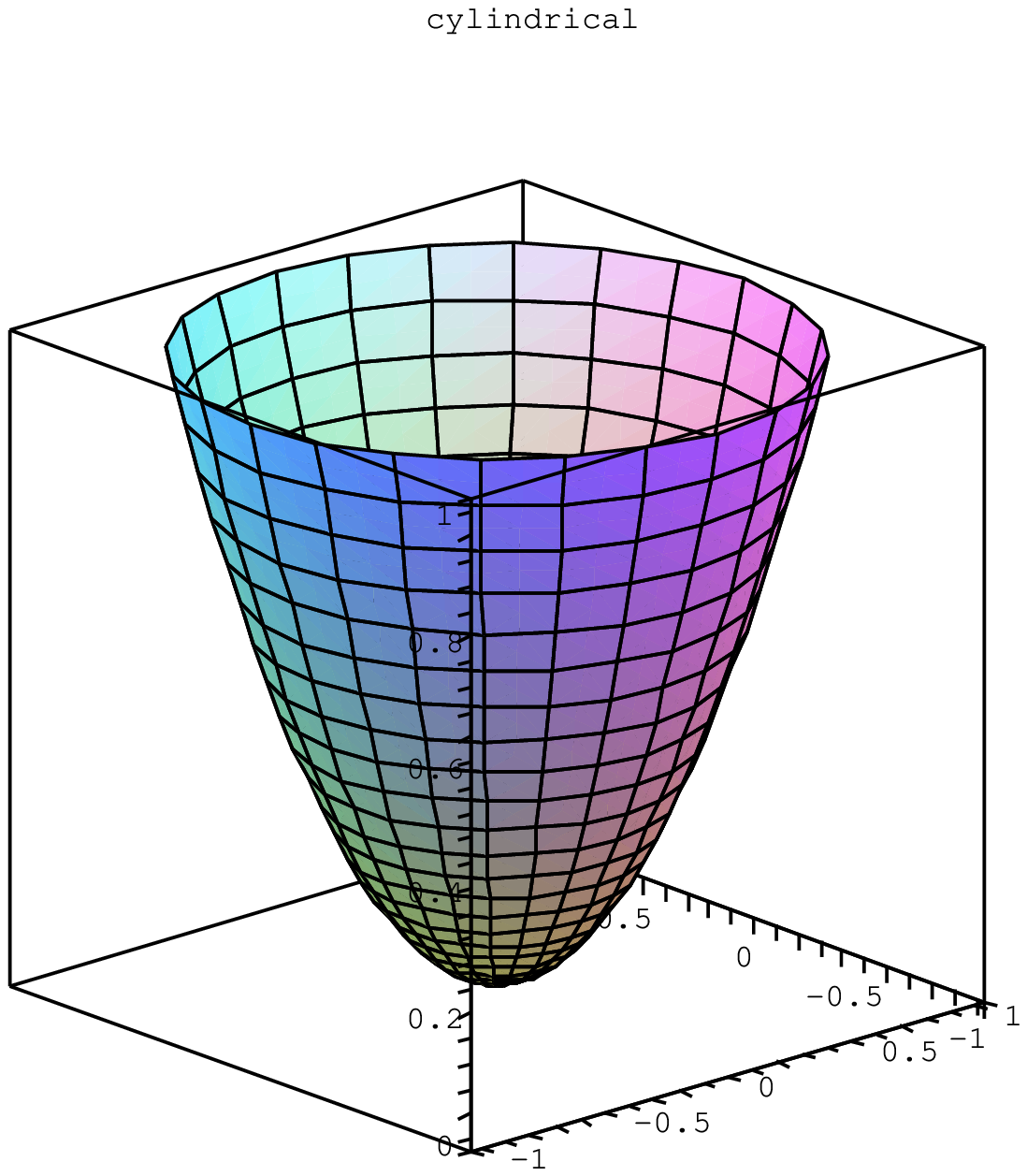,width=5cm}{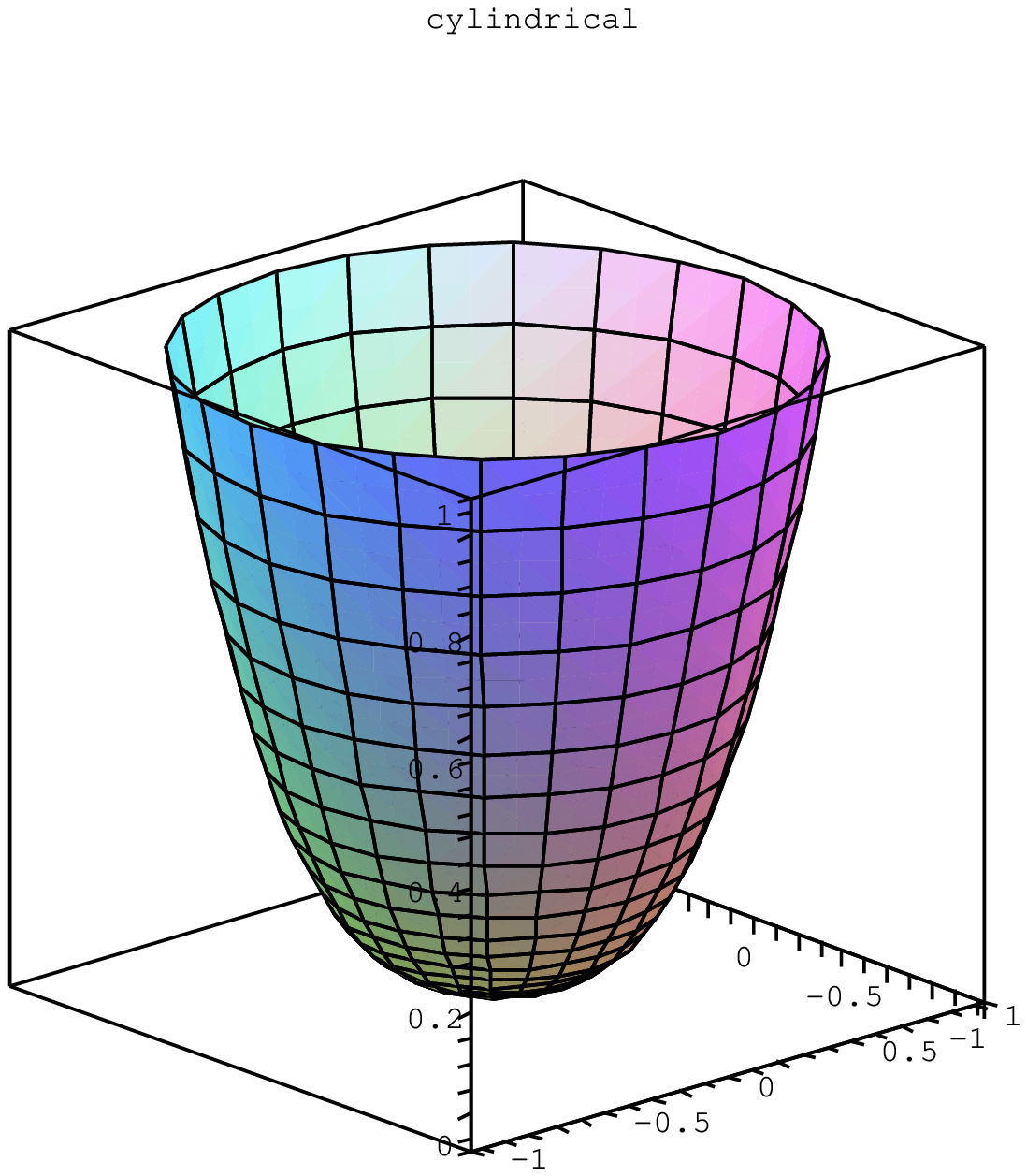,width=5cm}{Function $\rho^2(|z_1|^2,|z_2|^2)$ for the Bergmann manifold.}{$\rho^2(|z_1|^2,|z_2|^2)$ for Gutowski-Reall black hole.}
\EPSFIGURE[r]{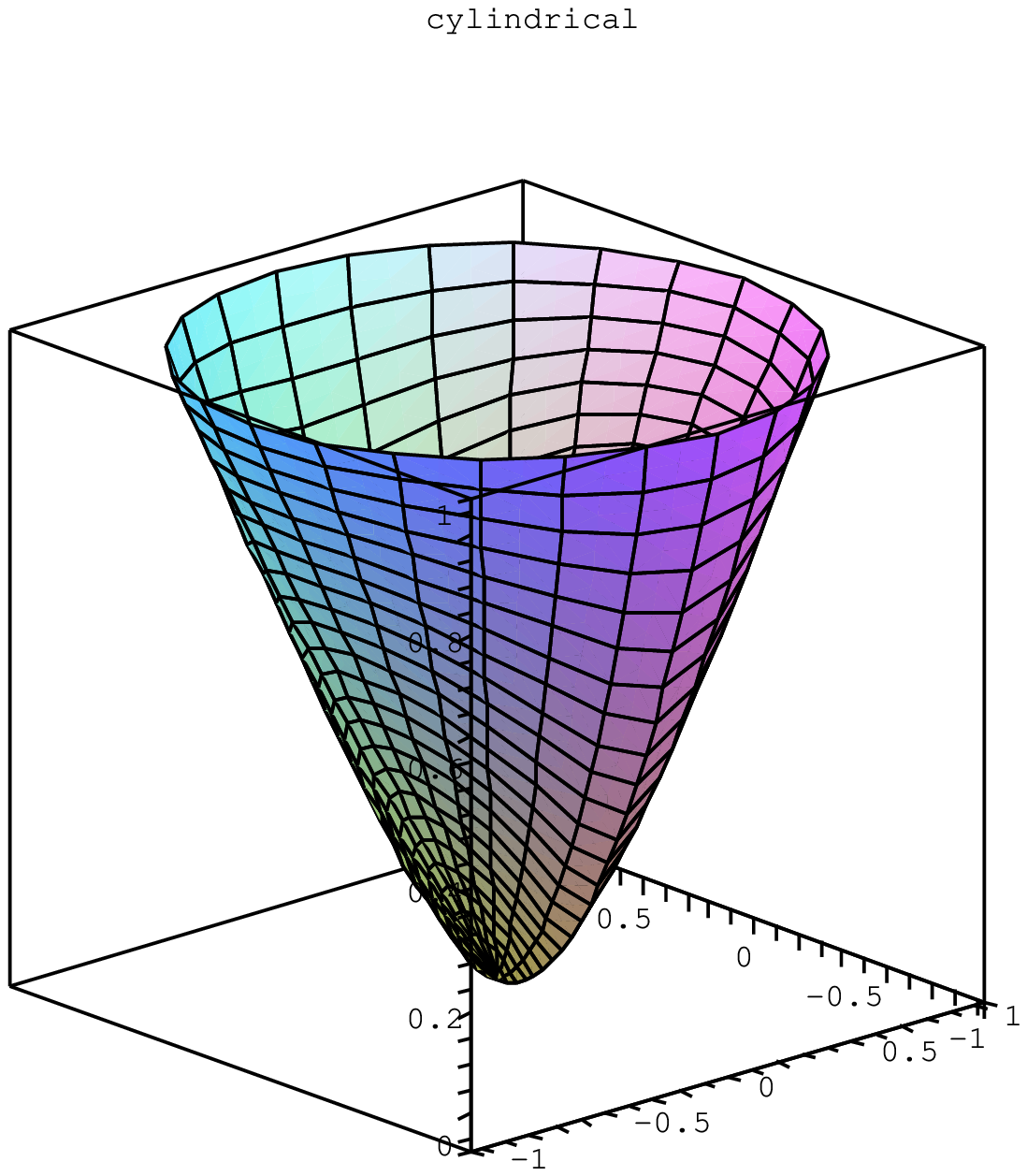,width=5cm}{$\rho^2(|z_1|^2,|z_2|^2)$ for Chong et al. black hole.}

%\begin{figure}[h!]
%\label{bergm}
%\centering\epsfig{file=bergmann.eps,width=6cm}
%\centering\epsfig{file=gutreall.eps,width=6cm}
%\centering\epsfig{file=pope.eps,width=6cm}
%\small{\caption{Function $\rho^2(|z_1|^2,|z_2|^2)$ for the Bergmann manifold, Gutowski-Reall and Chong et al. black holes, respectively. $\rho$ varies from $0$ to $1$. For the first two cases $\rho=const$ surfaces are circles, but for the third case  they are ellipses, which, however become a circle as $\rho\rightarrow 1$. This is a manifestation of the asymptotic $AdS_5$ structure of the most general black holes.}}
%\end{figure}

To prove the above claim, we make a coordinate transformation $(x^1,x^2)\rightarrow (\sigma,\theta)$ to the coordinates used in \cite{Kunduri:2006ek}:
\be
                  e^{2x^1}=\left(\tanh\,g\sigma\right)^{2A_1^2}\frac{A_2^2\sin^2\theta}{A_1^2\cos^2\theta+A_2^2\sin^2\theta} \ , \label{x1}
\ee
\be
                  e^{2x^2}=\left(\tanh\,g\sigma\right)^{2A_2^2}\frac{A_1^2\cos^2\theta}{A_1^2\cos^2\theta+A_2^2\sin^2\theta}\ . \label{x2}
\ee
From \eqref{eq:BH_curve} we then see that our function $\rho$ is related to the coordinate $\sigma$ used in \cite{Kunduri:2006ek} by $g\rho=\tanh\,g\sigma$ and, from (\ref{eq:def_of_rho}), we find that the function $G$ is, in terms of $(\sigma,\theta)$ coordinates,
\be
                 G=\frac{1}{4g^2}\ln\cosh^2g\sigma\ .
\ee
It can be now explicitly checked, using (\ref{metricGij}), that this potential generates the base space used in \cite{Kunduri:2006ek}, which is
\be\begin{split}
                ds^2_B=& d\sigma^2+\frac{\sinh^2g\sigma}{g^2}\left(\frac{d\theta^2}{\Delta_\theta}+\Delta_\theta\sin^2\theta\,\cos^2\theta\left[\frac{d\phi^1}{A_1^2}-\frac{d\phi^2}{A_2^2}\right]^2\right)\\
		       & \hspace{48pt}+\frac{\sinh^2g\sigma\cosh^2g\sigma}{g^2}\left[\frac{\sin^2\theta}{A_1^2}d\phi^1+\frac{\cos^2\theta}{A_2^2}d\phi^2\right]^2,
\end{split}\ee
with
\be 
\Delta_{\theta}=A_1^2\cos^2\theta+A_2^2\sin^2\theta \ . \label{deltapope}\ee

\section{The conical structure of the Black Hole solutions}\label{conical}
In the following we make some remarks on the structure of the K\"ahler base $\mcK$ near a possible horizon. We will present some arguments for $\mcK$ being a K\"ahler cone in the near horizon limit, for a certain class of solutions which includes all known black holes. We will state our assumptions as we exhibit the argument. 

\begin{description}
\item[$1)$] Near the horizon we assume that $f^{-1}$ should behave as
\be\label{eq:hypothesis}
                      f^{-1}=\frac{c(\theta)}{\rho^2} \ ,  
\ee    
where the coordinate $\rho$ vanishes at the horizon. We allow the numerator to depend on some \textit{other} coordinate, $\theta$. Note that the coordinates $(\rho,\theta)$ are functions of $(x^1,x^2)$. From (\ref{fricci}), we conclude that the base space will therefore have a curvature singularity at $\rho=0$. Note that $\rho$ in equation \eqref{eq:hypothesis}, is going to be the same as the $\rho$ in \eqref{eq:def_of_rho} up to the transformation explained after equation \eqref{after}.

\item[$2)$] In view of the singularity of the base, we impose that $f^{-1}ds_B^2(\mcK)$ has a finite proper size along the angular directions $\phi^i$, in the $\rho=0$ limit. Writing the metric in the form (\ref{metricGij}) and using $(\rho,\theta)$ coordinates instead or $(x^1,x^2)$, this means that the metric coefficients for the  $\phi$ coordinates should scale as $\rho^2$:
\be
           ds_B^2(\mcK)\simeq ds^2_2(\rho,\theta) + \rho^2 f_{ij}(\theta)\,d\phi^{i}d\phi^{j}.      \label{nonsing}
\ee 
The idea underlying this assumption is to allow a finite size horizon. However, one could conceive achieving a finite size horizon not by requiring  $f^{-1}ds_B^2(\mcK)$ to have a finite proper size along the angular directions, but rather by requiring the \textit{full metric} (\ref{timelikesolution}) to have a finite size along such directions. We will comment in subsection \ref{blackring} that, in principle, this is exactly what happens for a possible asymptotically $AdS_5$ black ring, which will, therefore, be excluded from our analysis.
\end{description}
Our assumptions and the equations (\ref{eq:hypothesis}) and (\ref{nonsing}) that followed, allow us to infer some more properties of the near-horizon geometry, from the way different objects transform under the following scaling: 
\[ \rho\to\lambda\rho \ ,  \ \ \ \ \ \theta\to\theta \ . \]
From (\ref{nonsing}) and (\ref{metricGij}) we see that 
\be G_{ij}\to\lambda^2G_{ij} \ , \label{scalinggij}\ee 
while from (\ref{eq:hypothesis}) and (\ref{fricci}), we see that 
\[ R\to R/\lambda^2 \ , \]
which are the behaviours to be expected from the proper size of angular coordinates and curvatures, respectively. It follows that $G^{ij}\to G^{ij}/\lambda^2$ and consequently, using (\ref{riccigij}), 
\be\label{eq:transf_x}
                 x^i\to x^i+f^{(i)}(\lambda) \ ,
\ee
where $f^{(i)}(\lambda)$, $i=1,2$, are arbitrary functions. Note that this is exactly what happens for the known black holes, as can be seen from (\ref{x1})-(\ref{x2}). Together with (\ref{scalinggij}) this last equation implies that the ``potential'' transforms as
\be\label{eq:transf_G}
                 G(x)\to \lambda^2 G(x) + f^{(3)}(\lambda) \ .
\ee
From this equation, we conclude that in terms of the $(\rho,\theta)$ coordinates, the ``potential'' $G$ must be proportional to $\rho^2$, up to an additive constant which is pure K\"ahler gauge. Thus,
\be
G(\rho,\theta)=\chi(\theta)\rho^2 \ . \label{after} \ee
We will redefine $\rho$ in such a way as to have $\chi(\theta)=1/4$. This amounts to a redefinition of the function $c(\theta)$ in eq.\eqref{eq:hypothesis}. We thus have
\be
                     G=\frac{\rho^2}{4}\ .
\ee
Note that  this coincides with the definition of $\rho$ given by equation \eqref{eq:def_of_rho} in the limit of small $\rho^2$. Therefore, we might suspect that all K\"ahler bases obeying our requirements are described by a two dimensional surface similar to (\ref{eq:BH_curve}). To see this is indeed the case note that under two consecutive re-scalings, (\ref{eq:transf_x}) implies that
\be
f^{(i)}(\lambda_1\lambda_2)=f^{(i)}(\lambda_1)+f^{(i)}(\lambda_2) \ . \ee
Therefore
\be f^{(i)}(\lambda)=A_i^2\ln \lambda \ , \ee
$i=1,2$, and we take $A_i^2\geq 0$, which can always be achieved by a holomorphic coordinate change, $z^i\to 1/z^i$. Thus, the combination
\[ t^i\equiv \frac{2x^i}{A_i^2}-2\ln\rho \ , \]  
is inert under scalings and therefore in $(\rho,\theta)$ coordinates $t^i=t^i(\theta)$ is a function of $\theta$ only. The function $\rho$ (and hence $G$) can thus be defined as a $2d$ surface in ${\mathbb R}^3$,
\be\label{eq:def_curve}
                    F(t^1,t^2)=0 \ ,
\ee
which is scaling independent. The case of all known black holes \eqref{eq:BH_curve} is given by the curve $F=e^{A_1^2t^1}+e^{A_2^2t^2}-1$, even away from the horizon.

We shall now determine the family of K\"ahler bases which obey our requirements together with one last technical assumption:
\begin{description}
\item[$3)$] We impose the convexity of the ``potential'' $G$. Since at $x^i=-\infty$, $G_{ij}=G_k=0$, it follows that (at least near the horizon) we have
\be
                        \frac{\partial G}{\partial x^i}\geq 0 \ .
\ee
A small computation, using (\ref{eq:def_curve}), then shows that
\be
                        \frac{\partial G}{\partial x^i}=\frac{\rho^2 \gamma_i(\theta)}{2} \ , \quad\textup{where}\quad \gamma_i=\frac{1}{A_i^2}\frac{\partial F}{\partial t^i}\frac{1}{\sum_i \frac{\partial F}{\partial t^i} } \ .
\ee
This expression makes clear that $\gamma_i\geq 0$ and that $\sum A_i^2 \gamma_i=1$. It is thus sensible to define the  $\theta$ coordinate through
\be
             \gamma_1=\frac{\sin^2\theta}{A^2_1},\quad \gamma_2=\frac{\cos^2\theta}{A_2^2}\ .
\ee
\end{description}
We now have all the information required to compute the generic form of the K\"ahler bases consistent with our assumptions, which we shall do in $(\rho, \theta, \phi^1,\phi^2)$ coordinates. The $g_{\rho\rho}$, $g_{\theta\rho}$ and $g_{\theta\theta}$ components are 
\be
g_{\rho\rho}=\frac{\partial x^i}{\partial \rho}\frac{\partial x^j}{\partial \rho}G_{ij}=\frac{\partial^2G}{\partial \rho^2}-\frac{\partial^2x^i}{\partial \rho^2}G_i=1 \ , \ee
\be
g_{\theta\rho}=\frac{\partial^2G}{\partial \rho\partial \theta}-\frac{\partial^2x^i}{\partial \rho \partial \theta}G_i=0 \ . \ee
\be
g_{\theta\theta}=-\frac{\partial^2x^i}{\partial \theta^2}G_i\equiv \frac{\rho^2}{\Delta_{\theta}} \ , \ee
which defines an arbitrary function of $\theta$, $\Delta_{\theta}$. This arbitrary function can also be re-expressed as
\be
                \Delta_{\theta}=\frac{2A_1^2}{\sin2\theta}\frac{\partial\theta}{\partial x^1} \ . 
\ee
We can now conclude that our K\"ahler base is a cone:
\be
                        ds_B^2= d\rho^2+\rho^2\,ds_3^2 \ ,
\label{eq:a_cone}
\ee
where the metric of the $d=3$ \emph{Sasakian} space is, using also (\ref{nonsing})
\be
                        ds_3^2=\frac{d\theta^2}{\Delta_{\theta}}+f_{ij}(\theta)\,d\phi^id\phi^j \ .
\ee 
The 2-dimensional metric $f_{ij}(\theta)$ is now determined by
\be
          f_{ij}(\theta)=\frac{G_{ij}}{\rho^2}=\gamma_i(\theta)\gamma_j(\theta)+\half\frac{\partial \gamma_i}{\partial\theta}\frac{\partial\theta}{\partial x^j}\ .
\ee
Noting that the symmetry between $i$ and $j$ determines that $\sum_i A_i^2\partial_i\theta=0$, we find that 
\be
               f_{ij}(\theta)d\phi^id\phi^j=\left[\frac{\sin^2\theta}{A_1^2}d\phi^1+\frac{\cos^2\theta}{A_2^2}d\phi^2\right]^2+\sin\theta\,\cos\theta\frac{\partial\theta}{\partial x^1}A_1^2\left[\frac{d\phi^1}{A_1^2}-\frac{d\phi^2}{A_2^2}\right]^2\ .
\ee
Thus, we can finally write our K\"ahler base as
\be
               ds_{B}^2=d\rho^2+\rho^2\left\{\frac{d\theta^2}{\Delta_{\theta}}+\left[\frac{\sin^2\theta}{A_1^2}d\phi^1+\frac{\cos^2\theta}{A_2^2}d\phi^2\right]^2+\Delta_{\theta}\sin^2\theta\,\cos^2\theta\left[\frac{d\phi^1}{A_1^2}-\frac{d\phi^2}{A_2^2}\right]^2\right\} \ . \label{basedelta1}
\ee
Our family of K\"ahler bases is fully determined up to just \emph{one arbitrary function} of $\theta$, $\Delta_{\theta}$. In the known black hole solutions this function is 
\be 
\Delta_{\theta}=A_1^2\cos^2\theta+A_2^2\sin^2\theta \ . \ee

\subsection{Sasakian space as a fibration over K\"ahler manifold}\label{fibration}
Taking the result from the previous section that the near horizon geometry of the base is a K\"ahler cone, we can obtain the metric \eqref{basedelta1} in a different way, which naturally introduces a very useful new coordinate.

It is a general result (see e.g.
\cite{Martelli:2006yb}) that a $(2n+1)$-dimensional Sasakian space can be
written as a fibration over a $2n$-dimensional K\"ahler space. In the case at
hand $n=1$ and the base is two dimensional. To see this we note that the
K\"ahler potential of a four dimensional K\"ahler cone can be written as
\be
                   \mcK=|z|^2e^{2{\hat \mcK}(w,{\bar w})}\equiv \frac{1}{2}\rho^2 \ .
\ee
A tedious calculation then shows that
\be\label{eq:sasakian2}
                     ds_3^2=\eta\otimes\eta+2{\hat \mcK}_{w{\bar w}}dw\,d{\bar w} \
,
\ee
where the 1-form $\eta$ is
\be
                    \eta=d\phi-i({\hat \mcK}_{w}dw-{\hat \mcK}_{\bar w}d{\bar w}) \ ,
\ee
and $\phi$ is defined through $z=\rho e^{-\hat{\mcK}}e^{i\phi}/\sqrt{2}$. This also shows that the K\"ahler form $J_2$ of the two dimensional base is
related to $\eta$ as
\be
                     J_2=-\half d\eta \ .
\ee
Furthermore, for the K\"ahler cone one finds that
\be
                     J=-\half d[\rho^2\eta] \ .
\ee
Two additional suggestive relations between the geometry of the four dimensional
K\"ahler cone and the two dimensional base are
\be
                    \mcR=4J_2+\mcR_2 \ ,
\ee
and
\be
                     R=\frac{R_2-8}{\rho^2} \ ,
\ee
where $\mcR_2$ and $R_2$ are the Ricci form and the Ricci scalar of the 2d base.

From the present perspective, the cones we presented above are obtained by
requiring the 2d base space to be toric. We comment on non-toric cones in appendix \ref{nontoric}. Being toric means (see section \ref{seckahler}) that the metric
of the base can be written as
\be
                     ds_2^2=G_2''(y)[dy^2+d{\bar\psi}^2] \ ,
\ee
where $2G_2(y)$ is the K\"ahler potential written in suitable coordinates, while
$J_2$ is
\be
                     J_2=-d(G_2'd{\bar\psi}) \ .
\ee
We can change to symplectic coordinates $(x,\psi)$ defined through
\be
                     x=2G'_2(y),\quad \psi=2{\bar\psi} \ .
\ee
With $H(x)\equiv G_2''(y(x))$, we rewrite the metric as
\be
                    ds_2^2=\frac{dx^2}{4H(x)}+\frac{1}{4}H(x)d\psi^2 \ ,
\label{2dbase}
\ee
and the K\"ahler form as
\be
                      J_2=-\frac{1}{4}d(xd\psi) \ .
\ee
This then shows that $2\eta=d\phi+x d\psi$ and the 3d Sasakian space is
\be
ds_3^2=\frac{dx^2}{4H(x)}+\frac{1}{4}H(x)d\psi^2+\frac{1}{4}(d\phi+x d\psi)^2 \
.
\label{eqn:3dsasaki}
\ee
Setting 
\be
x=\cos2\theta \ ,  \ \ \ \ \ \ \frac{H(x)}{1-x^2}=\Delta_\theta \ , \label{xh}
\ee
\[ \psi=\frac{\phi^2}{A_2^2}-\frac{\phi^1}{A_1^2} \ , \ \ \ \ \ \ 
\phi=\frac{\phi^2}{A_2^2}+\frac{\phi^1}{A_1^2} \ , \] 
and using \eqref{eq:a_cone} we retrieve the cone \eqref{basedelta1}. Note that $H(x)$ can be faced as the conformal factor of the two dimensional space (\ref{2dbase}).

\subsection{A comment on supersymmetric black rings}\label{blackring}
Consider the asymptotically flat supersymmetric black ring found in \cite{Elvang:2004rt}. The metric still has the form \eqref{timelikesolution}, but where the base space is now a hyper-K\"ahler manifold, in fact flat space $\mathbb{R}^4$ conveniently written in ring type coordinates:
\be
ds^2_ B(\mathbb{R}^4) = \frac{R^2}{(x-y)^2}\left[ \frac{dy^2}{y^2-1} +
(y^2-1)d\psi^2+\frac{dx^2}{1-x^2}+(1-x^2)d\phi^2 \right] \ .
\label{base}
\ee
The remaining metric functions are
\be
f^{-1}=1+\frac{Q-q^2}{2R^2}(x-y)-\frac{q^2}{4 R^2}(x^2-y^2) \ , 
\ee
and $\omega= \omega_{\psi}(x,y) d\psi+\omega_{\phi}(x,y) d\phi$, where
\beqa
\label{omegas}
\omega_\phi &=& -\frac{q}{8R^2} (1-x^2) \left[3Q - q^2 ( 3+x+y)
 \right]\,, \\
\omega_\psi &=& \frac{3}{2} q(1+y)  + \frac{q}{8R^2} (1-y^2)
\left[3Q - q^2 (3 +x+y) \right]
\,.\nonumber
\eeqa
The solution has three independent parameters: $R$, $Q$ and $q$ which are all positive constants; the latter two are proportional to the net charge and
to the local dipole charge of the ring, respectively. The coordinate ranges are $-1\le x\le 1$, $-\infty\le y\le -1$, $0\le \phi,\psi\le 2\pi$, and the near horizon limit of this solution is obtained by taking $y\rightarrow -\infty$. Taking this limit and introducing simultaneously the coordinates $\rho,\theta$ as $\rho=-R/y$ and $\theta=\arccos x$, we have 
\be
f^{-1}=\frac{q^2}{4\rho^2}+\mathcal{O}\left(\frac{1}{\rho}\right) \ ,  \label{flatring} \ee
and 
\be
ds^2_B(\mathbb{R}^4) \simeq d\rho^2 +\rho^2d\theta^2+R^2d\psi^2+\rho^2\sin^2\theta d\phi^2 \ .\label{basestring}
\ee

It is simple to see that this ring solution violates the assumptions we have made above to derive \eqref{basedelta1}. Indeed, although assumption $1)$ is obeyed, as can be checked by comparing \eqref{flatring} with \eqref{eq:hypothesis}, assumption $2)$ is clearly violated, since $\psi$ and $\phi$ are the $\phi^i$ coordinates of the previous section and the base space \eqref{basestring} does not fall in the form \eqref{nonsing}, due to the scaling of the metric coefficient for  $\psi$, $g_{\psi\psi}^B$. This coefficient is constant and therefore $f^{-1}g_{\psi\psi}^B$ diverges on the horizon. So, $\rho=0$  represents, in the base space, a circle of radius $R$, which becomes a singular surface for the metric $f^{-1}ds^2_B$, but becomes again a regular surface, in fact a ring shaped horizon, in the full metric. This is made possible by a cancellation of the divergent terms coming from both $f^{-1}ds^2_B$ and $f^2\omega_{\psi}^2$. Explicitly, expanding the metric functions to second order
\be 
f^{-1}=\frac{q^2}{4\rho^2}\left(1+\frac{2(Q-q^2)}{q^2R}\rho+\mathcal{O}\left(\rho^2\right)\right) \ , \ \ \ \ \omega_{\psi}=-\frac{q^3R}{8\rho^3}-\frac{q}{8\rho^2}[3Q-(3+\sin\theta)q^2]+\mathcal{O}\left(\frac{1}{\rho}\right) \ , \ee
as well as the metric coefficient $g_{\psi\psi}^B$ of the base space
\be
g_{\psi\psi}^B=R^2-2\sin\theta R\rho+\mathcal{O}(\rho^2) \ , \ee
we find that
\be
g_{\psi\psi}=-f^2\omega_{\psi}^2+f^{-1}g_{\psi\psi}^B=\mathcal{O}(\rho^0) \ . 
\ee
Indeed, the divergent terms ($\mathcal{O}(1/\rho)$ and $\mathcal{O}(1/\rho^2)$) all cancel leaving only finite contributions to the proper size of the $\psi$ direction.

We expect therefore, that the base space of a possible $AdS_5$ black ring will not approach a K\"ahler cone near the horizon.

\section{5D Near horizon solutions from K\"ahler cones} 
Our goal is now to determine what are the constraints on $\Delta_{\theta}$ such that, given the base \eqref{basedelta1}, a non-trivial five dimensional solution exists. Even though it has been claimed \cite{Gauntlett:2003fk,Gutowski:2004ez} that a five dimensional solution always exists for any four dimensional K\"ahler base, we will show that this is not the case. In this section we obtain a constraint equation on $\Delta_{\theta}$, choosing a special ansatz for the 2-form $G^-$ and hence the five dimensional solution. In the next section we will show that this constraint still holds even if we take the most generic form for the five dimensional solution.

The Ricci scalar for (\ref{basedelta1}) is 
\be
              R=-\frac{1}{\rho^2}\left[8(1-\Delta_{\theta})+\frac{1}{\sin^32\theta}\frac{d}{d\theta}\left(\sin^32\theta \frac{d\Delta_{\theta}}{d\theta}\right) \right]\equiv -\frac{24g^2 }{\rho^2}c(\theta) \ , \label{rc} \ee
which defines the function $c(\theta)$ and determines the metric function $f$, via (\ref{fricci}), to be 
\be f^{-1}=\frac{c(\theta)}{\rho^2} \ . \label{fc}\ee 
The Ricci form \eqref{ricciformg}  can  be found to be
\be
              {\mcR}_{ij}=\frac{R}{2}\left(G_{ij}-\frac{4}{\rho^2}G_iG_j\right)=\frac{R}{2}\,\rho^2\Delta_{\theta}\sin^2\theta\,\cos^2\theta\,\frac{(-1)^{i+j}}{A_i^2A_j^2}\ .
\ee
This can then be used to determine $G^+$,
\be
                       G^+_{ij}=-\frac{R}{4g}\left(\half G_{ij} - \frac{4}{\rho^2}G_iG_j \right)\ ,
\ee
and $(G^+)^2$,
\be
                      G^{+ij}G^+_{ij}=\frac{R^2}{16g^2}\ .
\label{eq:Gsquared}
\ee
Plugging this back in \eqref{eq:to_solve1} we find that 
\be
              \nabla^2 f^{-1}=-gf^{-1}G^{-mn}J_{mn}\ .
\label{eqreduzida}
\ee
Recall that according to our assumptions, this equation is valid only at the horizon; moreover, it only determines the components of $G^-$ `parallel' to the K\"ahler form; thus we will have some undetermined functions in the general form allowed by this equation for $G^-$. To write such an ansatz for $G^-$, it is useful to introduce the following set of 1-forms
\be\nonumber
                 {\bf e}^{\rho}=d\rho \ ,\ \ \ \ \quad {\bf e}^\theta=\rho\, \frac{d\theta}{\sqrt{\Delta_\theta}}\ ,
\ee
\be
                 {\bf e}^1=\rho\left[\frac{\sin^2\theta}{A_1^2}d\phi^1+\frac{\cos^2\theta}{A_2^2}d\phi^2\right] \ , \ \ \ \ 
  {\bf e}^2=\rho \sqrt{\Delta_\theta}\sin\theta\,\cos\theta\left[\frac{d\phi^1}{A_1^2}-\frac{d\phi^2}{A_2^2}\right]\ .
\ee
In this basis, the K\"ahler form and $G^+$ read
\be
                 J=-\left({\bf e}^\rho\wedge {\bf e}^1+{\bf e}^\theta\wedge {\bf e}^2\right)\ ,
\ee
\be
                 G^+=-3gf^{-1}\left({\bf e}^\rho\wedge {\bf e}^1-{\bf e}^\theta\wedge {\bf e}^2\right)\ .
\label{Gplusformas}
\ee
Furthermore, the 4d Hodge star product is chosen to act as $\star ({\bf e}^\rho\wedge {\bf e}^1)=-{\bf e}^\theta\wedge {\bf e}^2$ and $\star ({\bf e}^\rho\wedge {\bf e}^2)={\bf e}^\theta\wedge {\bf e}^1$.
The most general form for $G^-$ (in leading order in $\rho$) that allows the symmetries of the base space to be present in the five dimensional spacetime is 
\be
G^-=\frac{l(\theta)}{\rho^2}({\bf e}^{\rho}\wedge {\bf e}^1+{\bf e}^{\theta}\wedge {\bf e}^2)+\frac{n(\theta)}{\rho^2}({\bf e}^{\rho}\wedge {\bf e}^2-{\bf e}^{\theta}\wedge {\bf e}^1)+\frac{m(\theta)}{\rho^b}({\bf e}^1\wedge {\bf e}^2-{\bf e}^{\theta}\wedge {\bf e}^{\rho}) \ ; \label{Gminusformas}\ee
we will come back to this point in section \ref{fullspa}. It follows that 
\be
               G^{-nm}J_{nm}=2\star\left(\star G^-\wedge J\right)=-\frac{4l(\theta)}{\rho^2}\ .  
\ee
From \eqref{eqreduzida} and (\ref{fc}) we can determine $l(\theta)$ in terms of the derivatives of $c(\theta)$, 
\be\label{eq:defl}
              l(\theta)=\frac{1}{4g\,c(\theta)\sin\theta\cos\theta}\frac{d}{d\theta}\left(\Delta_\theta\sin\theta\cos\theta\frac{dc(\theta)}{d\theta}\right), 
\ee 
where $c(\theta)$ was defined in (\ref{rc}) to be
\be
c(\theta)=\frac{1}{24g^2}\left[8(1-\Delta_{\theta})+\frac{1}{\sin^32\theta}\frac{d}{d\theta}\left(\sin^32\theta \frac{d\Delta_{\theta}}{d\theta}\right) \right] \ . \label{eqn:defc} \ee
Combining the expressions for $f$, $G^+$ and $G^-$ given by (\ref{fc}), (\ref{Gplusformas}) and (\ref{Gminusformas}) we compute $dw$ via 
\begin{equation}
dw=f^{-1}(G^+ +G^-) \ ; 
\label{eqn:dw2}
\end{equation}
the requirement that this form is closed (\ref{closed}) determines $b=0$ and the following three conditions:
\be
m(\theta)=\frac{constant}{\sin\theta\cos\theta\sqrt{\Delta_{\theta}}c(\theta)} \ , \ee
\be
              n(\theta)=\frac{\sqrt{\Delta_\theta}}{2c(\theta)}\frac{d}{d\theta}\left[c(\theta)\left(l(\theta)-3g\,c(\theta)\right)\right]\ , 
\ee
so that both $m(\theta)$ and $n(\theta)$ are completely determined in terms of $c(\theta)$; finally
\be
            \frac{d}{d\theta}\left(\Delta_\theta\sin\theta\cos\theta\frac{d}{d\theta}\left[c(\theta)\left(l(\theta)-3g\,c(\theta)+\frac{2}{g}\right)\right]\right)=0 \ . \label{eq6ordem}
\ee
This is a $6^{th}$ order ODE on $\Delta_{\theta}$ which is considerably involved. But it can be reduced to a much simpler form in the following way. Introduce new independent and dependent variables $x$ and $H(x)$, as in \eqref{xh}. Then we have 
\be
c(x)=\frac{2+H''}{6g^2} \ , \ \ \ \ \ l(x)=\frac{1}{g(2+H'')}\left(HH'''\right)' \ , \ee
where 'primes' denote $x$ derivatives. After some manipulations (\ref{eq6ordem}) becomes the remarkably simpler equation
\be
\left(H^2H''''\right)''=0 \ . \label{hequation}\ee
This equation can be now reduced to a fourth order equation:
\be
H^2H''''=\alpha x+\beta \ . \label{hequationab}\ee
For $\alpha=\beta=0$, the general solution is a cubic polynomial in $x$:
\be
H(x)=a_0+a_1x+a_2x^2+a_3x^3 \ . \label{cubicpol}\ee
All known black holes fall in this class of solutions. We will analyse the solutions of \eqref{hequation} in section \ref{secsolutions}, after showing, in section \ref{fullspa}, that this equation, which was obtained from a near horizon analysis, still arises in a full spacetime analysis. Moreover, we will show in section \ref{fullspa} that, even taking the most general form for $G^-$, the constraint \eqref{hequation} still holds. Thus, as claimed above, not all K\"ahler bases will provide a non-trivial five dimensional solution.

\section{Full spacetime analysis}\label{fullspa}
As we are searching for asymptotically AdS$_5$ BH solutions, it is sensible to find K\"ahler bases which for $\rho\to 0$ reduce to the above discussed cones while for $\rho \to 1$ approach Bergmann space. It is not difficult to find that
\be\label{eq:general_G}
                   G=-\frac{1}{4g^2}\ln\left(1-g^2\rho^2\right),
\ee 
where $\rho$ is defined exactly as in \eqref{eq:def_curve}, fulfils these criteria, as we will now see. Note that one could discuss forms  of $G=G(\rho^2)$ other than \eqref{eq:general_G} but we will not pursue such analysis in this paper.\footnote{A class of K\"ahler-Einstein spaces that differ from Bergmann space in the form of the function $G(\rho)$ was studied in  \cite{Gauntlett:2004hh}.}

 We easily find that
\be
                   G_i=\frac{1}{2}\frac{\rho^2}{1-g^2\rho^2}\gamma_i(\theta),
\ee
and
\be
                   G_{ij}=\frac{\rho^2}{1-g^2\rho^2}\left[\frac{\gamma_i(\theta)\gamma_j(\theta)}{1-g^2\rho^2}+\half\frac{\partial \gamma_i}{\partial\theta}\frac{\partial\theta}{\partial x^j}\right].
\ee
With $\rho^2=\tanh^2g\sigma/g^2$, we can now write the metric of the full K\"ahler base as
\be\begin{split}
                ds^2_B=& d\sigma^2+\frac{\sinh^2g\sigma}{g^2}\left(\frac{d\theta^2}{\Delta_\theta}+\Delta_\theta\sin^2\theta\,\cos^2\theta\left[\frac{d\phi^1}{A_1^2}-\frac{d\phi^2}{A_2^2}\right]^2\right)\\
		       & \hspace{48pt}+\frac{\sinh^2g\sigma\cosh^2g\sigma}{g^2}\left[\frac{\sin^2\theta}{A_1^2}d\phi^1+\frac{\cos^2\theta}{A_2^2}d\phi^2\right]^2, \label{basefull}
\end{split}\ee
while the K\"ahler 2-form reads
\be
                    J=-d\left[\frac{\sinh^2g\sigma}{2g^2}\left(\frac{\sin^2\theta\,d\phi^1}{A_1^2}+\frac{\cos^2\theta\,d\phi^2}{A_2^2}\right)\right].
\ee
Notice that these spaces include the K\"ahler bases discussed in \cite{Kunduri:2006ek} (which correspond to take $\Delta_{\theta}$ given by \eqref{deltapope}). Remarkably, we obtained them from making some mild assumptions on the behaviour of the metric near a singularity. Finally, we record the Ricci form for later convenience:
\begin{equation}
{\cal R}=-\frac{3}{2}d\left\{
	\sinh^2g\sigma(d\phi+xd\psi)+\frac{x}{3}\left[
		2-A_1^2-A_2^2-\frac{3}{2}(A_1^2-A_2^2)x\right]d\psi\right\}
\label{eqn:Ricciform}
\end{equation}

For $g\sigma\to 0$ the bases reduce to the above discussed K\"ahler cones as they should. To see that in the $g\sigma\to\infty$ limit these spaces have negative constant scalar curvature note that
\begin{equation}
R=-24g^2\left[1+\frac{g^2 c(\theta)}{\sinh^2g\sigma}\right]\;,
\label{eqn:fRicci}
\end{equation}
where $c(\theta)$ was defined in \eqref{rc}. 

We now show that a full spacetime analysis of the supersymmetry constraints leads to the same equation (\ref{hequation}) as the near horizon analysis. It is simpler to do the analysis in $x$ coordinates, in terms of which the K\"ahler base \eqref{basefull} can be written
\be ds^2_B=d\sigma^2+\frac{\sinh^2g\sigma}{4g^2}\left(\frac{dx^2}{H(x)}+H(x)d\psi^2+\cosh^2g\sigma(d\phi+xd\psi)^2\right) \ . \label{basex} \ee
This is a remarkably simple form for the base space. In particular notice that the known black holes simply correspond to taking $H(x)$ to be a cubic polynomial.  In terms of the obvious tetrad
\be
\begin{array}{c}
\displaystyle{{\bf e}^{\sigma}=d\sigma \ , \ \ \ \ \ \ \ \ {\bf e}^x=\frac{\sinh g\sigma}{2g}\frac{dx}{\sqrt{H}} \ ,} \vspace{0.4cm}\\ 
\displaystyle{{\bf e}^{\psi}=\frac{\sinh g\sigma}{2g}\sqrt{H}d\psi \ , \ \ \ \ \ \ \ \ {\bf e}^{\phi}=\frac{\sinh g\sigma\cosh g\sigma}{2g}(d\phi+xd\psi) \ ,} \end{array} \ee
the K\"ahler form and Ricci form are
\be
J=-({\bf e}^{\sigma}\wedge {\bf e}^{\phi}+{\bf e}^x\wedge {\bf e}^{\psi}) \ , \ee
\be
\mathcal{R}=12g^2(1-f^{-1}){\bf e}^x\wedge {\bf e}^{\psi}-6g^2({\bf e}^{\sigma}\wedge {\bf e}^{\phi}+{\bf e}^x\wedge {\bf e}^{\psi}) \ , \ee
whereas the function $f$, computed from the Ricci scalar is 
\be
f^{-1}=1+\frac{2+H''}{6\sinh^2g\sigma} \ . \ee
It is curious to notice that the K\"ahler form does not depend on $H(x)$ and hence, it is the same for the whole family of bases. The Ricci form, of course, depends on $H(x)$. From \eqref{Gplus} it follows that 
\be
G^+=3g(1-f^{-1})({\bf e}^{\sigma}\wedge{\bf e}^{\phi}-{\bf e}^x\wedge{\bf e}^{\psi}) \ . 
\label{eqn:Gplusx}
\ee
From \eqref{eq:to_solve1} we determine the part of $G^-$ parallel to $J$. Thus, the most general expression for $G^{-}$ is 
\be 
G^{-}=l(x,\sigma)({\bf e}^{\sigma}\wedge {\bf e}^{\phi}+{\bf e}^x\wedge {\bf e}^{\psi})+fN(\sigma,x,\phi,\psi)({\bf e}^{\sigma}\wedge {\bf e}^{\psi}-{\bf e}^x\wedge {\bf e}^{\phi})+fM(\sigma,x,\phi,\psi)({\bf e}^{\sigma}\wedge {\bf e}^{x}-{\bf e}^{\phi}\wedge {\bf e}^{\psi}) \ , 
\label{eqn:Gminusx}
\ee
where 
\be 
l(x,\sigma)= gf\left(2+\frac{2+H''}{2\sinh^2g\sigma}+\frac{(H'''H)'}{6\sinh^4g\sigma}\right) \ , \ee
and $N,M$ are arbitrary functions of their arguments. Since we are taking the most generic form of $G^-$, the restriction given by the integrability conditions on $H(x)$, if any, indeed restrict the K\"ahler bases that can generate five dimensional solutions. These integrability conditions can be computed from \eqref{closed} and yield the following four equations:
\be
\partial_{\phi}M=\frac{\partial_{\sigma}(N\cosh g\sigma \sinh^2 g\sigma)}{2g\sinh g\sigma}+\frac{g\sqrt{H} \cosh g\sigma }{6\sinh^4 g\sigma}\left[(H'''H)'-\frac{(2+H'')^2}{2}\right]' \ ; \label{int1} \ee
\be
\frac{(\partial_{\psi}-x\partial_{\phi})M}{\sqrt{H}}=\partial_x(N\sqrt{H})+\frac{g\cosh g\sigma}{3\sinh^4 g\sigma}(H'''H)' \ ; \label{int2} \ee
\be
\partial_{\phi}N=-\frac{\partial_{\sigma}(M\cosh g\sigma \sinh^2 g\sigma)}{2g\sinh g\sigma} \ ; \label{int3} \ee
\be
\frac{(\partial_{\psi}-x\partial_{\phi})N}{\sqrt{H}}=-\partial_x(M\sqrt{H}) \ . \ee
Define the new dependent variables
\be 
\tilde{M}(\sigma,x,\phi,\psi)\equiv \cosh g\sigma \sinh^2 g\sigma M(\sigma,x,\phi,\psi) \ , \label{Mtilde}\ee
\be 
\tilde{N}(\sigma,x,\phi,\psi)\equiv \cosh g\sigma \sinh^2 g\sigma N(\sigma,x,\phi,\psi)-\frac{g\sqrt{H}}{6\sinh^2 g\sigma}\left[(H'''H)'-\frac{(2+H'')^2}{2}\right]' \ , \label{Ntilde}\ee
and the new independent variable
\be u\equiv 4g\int \frac{d\sigma}{\sinh 2g\sigma}=-4{\rm arcth}\left(e^{-2g\sigma}\right) \ , \ee
where $u$ varies from $u=-\infty$ at a possible horizon and $u=0$ at spatial infinity; one finds, from \eqref{int1} and \eqref{int3}, that the function $\tilde{M}$ and $\tilde{N}$ are harmonic in the $u,\phi$ plane:
\be
\left(\partial_{\phi}^2+\partial_u^2\right)\tilde{M}=0 \ , \ \ \ \ \ \ \ \left(\partial_{\phi}^2+\partial_u^2\right)\tilde{N}=0 \ . \label{eqn:harmonic}\ee
The fact that we find harmonic equations is reminiscent of other ``linearisations'' that arise in supersymmetric solutions, namely the ones that allow multi black hole spacetimes and indicate ``no-force'' configurations. In this case, this linearisation indicates, as we shall see below,  the ability of superimposing in any background defined by a base with some $H(x)$, an infinite set of angular and radial deformations. It follows that
\be
\tilde{M}=\sum_{\alpha}c^M_{\alpha}(x,\psi)\mathcal{H}_{\alpha}(u,\phi) \ , \ \ \ \ \ \ \ \ \ \tilde{N}=\sum_{\alpha}c^N_{\alpha}(x,\psi)\mathcal{H}_{\alpha}(u,\phi) \ , \label{harmonicex} \ee
where $\{\mathcal{H}_{\alpha}\}$ is a basis of harmonic functions. Substituting \eqref{Mtilde} and \eqref{Ntilde} with \eqref{harmonicex} this expressions in (\ref{int2}), we find the integrability condition
\be
\left(H''''H^2\right)''=6\sinh^2g\sigma\left\{-\frac{(HH''')'}{3}+\frac{1}{g}\sum_{\alpha}\left[\frac{\partial_{\psi}c_{\alpha}^M-xc_{\alpha}^M\partial_{\phi}}{\sqrt{H}}-\partial_x (\sqrt{H}c_{\alpha}^N)\right]\mathcal{H}_{\alpha}\right\} \ . 
\label{eqn:int3}
\ee
The left hand side does not depend on either $\sigma$ or $\phi$; it depends only on $x$. Since $\mathcal{H}_{\alpha}$ is either a function of both $\sigma$  and $\phi$ or it is linear in one of them, the right hand side is never a function solely of $x$. Thus both sides must vanish. We therefore recover condition \eqref{hequation}, but we have the additional constraint:
\be 
\sum_{\alpha}\left[\frac{\partial_{\psi}c_{\alpha}^M-xc_{\alpha}^M\partial_{\phi}}{\sqrt{H}}-\partial_x (\sqrt{H}c_{\alpha}^N)\right]\mathcal{H}_{\alpha}=\frac{g(HH''')'}{3} \ . \label{const1} \ee
The remaining integrability conditions give the following constraints for the coefficients $c_{\alpha}^N$ and $c_{\alpha}^M$:
\be 
\sum_{\alpha}\left[\partial_{x}(\sqrt{H}c_{\alpha}^M)+\frac{\partial_{\psi}c_{\alpha}^N-xc_{\alpha}^N\partial_{\phi}}{\sqrt{H}}\right]\mathcal{H}_{\alpha}=0 \ ,\label{const2} \ee
\be \sum_{\alpha}\left(c_{\alpha}^M\partial_{\phi}-c_{\alpha}^N\partial_u\right)\mathcal{H}_{\alpha}=0 \ , \label{const3} \ee
\be \sum_{\alpha}\left(c_{\alpha}^M\partial_{u}+c_{\alpha}^N\partial_{\phi}\right)\mathcal{H}_{\alpha}=0 \ . \label{const4}\ee
Constraints \eqref{const1}-\eqref{const4}, together with \eqref{hequation} define the most general solution for our family of K\"ahler bases. To get it explicitly, we have to integrate the equation for $w$ \eqref{eqn:dw2}, where $G^+$ and $G^-$ are given in \eqref{eqn:Gplusx} and \eqref{eqn:Gminusx} respectively. As is well known, the field equations of the theory are all satisfied once we impose the equations of motion for the Maxwell field \cite{Gutowski:2004ez,Kunduri:2006ek}. Writing the field strength as
\begin{equation}
F=\frac{\sqrt 3}{2}d\left[f(dt+w)\right]+\frac{1}{2g\sqrt 3}~{\cal R}\;,
\end{equation}
and using \eqref{eqn:Ricciform}, it is clear that once we have obtained a solution for $w$, we can easily find potentials such that $F=dA$ and hence the Bianchi identities are trivially satisfied. It can be similarly checked that the dynamical Maxwell equations are satisfied and hence all supergravity equations.

\section{Solutions}\label{secsolutions}
As we mentioned before, the spatial isometries of the base need not be present in the full five dimensional solutions. We will split the analysis of the solutions into two cases: when the five dimensional solution has $U(1)^2$ spatial isometry and when it does not. 

\subsection{Solutions with $U(1)^2$ spatial isometry}
The simplest way to obey constraints \eqref{const1}-\eqref{const4} is to take
\be
 c_{0}^N=-\frac{g\sqrt{H}H'''}{3}-\frac{C_1}{\sqrt{H}} \ , 
\label{eqn:allbhs} 
\ee
and all remaining $c_{\alpha}^N$ and $c_{\alpha}^M$ zero, where $\mathcal{H}_0=1$ is the constant harmonic form. This ensures that $w$ does not depend on $\phi$ or $\psi$. It follows that we can determine $w$ explicitly even without solving the remaining integrability condition \eqref{hequation}. Start by noting that
\begin{eqnarray}
\left[(H'''H)'-\frac{(2+H'')^2}{2}\right]'&=&\frac{1}{H}\frac{d}{dx}\left(H^2 H''''\right)-2H'''\;,\nonumber\\
	&=&\frac{C_H}{H}-2H'''\;,
\end{eqnarray}
where, from \eqref{hequation}, we have defined $(H^2 H'''')'=C_H$. $C_H$ is a constant and vanishes for all known $AdS$ black holes, since they will be described by cubic polynomials.  If follows that $\tilde M=M=0$ and, from \eqref{Ntilde},
\begin{equation}
N=-\frac{1}{\cosh(g\sigma)\sinh^2(g\sigma)}\left[\frac{g\cosh^2(g\sigma)\sqrt H H'''}{3\sinh^2(g\sigma)}
	+\frac{1}{\sqrt H}\left(C_1-\frac{gC_H}{6\sinh^2(g\sigma)}\right)\right]\; \ .
\end{equation}
Then, using \eqref{eqn:dw2} we find
\begin{equation}
w=w_\phi(x,\sigma) {\bf e}^{\phi}+w_\psi(x,\sigma) {\bf e}^{\psi}\; \ , 
\end{equation}
with
\begin{equation}
\begin{aligned}
w_\phi&=\frac{1}{\sinh(2g\sigma)}\Bigg\{2\sinh^2(g\sigma)-\frac{1}{6\sinh^2(g\sigma)}
	\left[(H'''H)'-\frac{(2+H'')^2}{2}\right]+\frac{1}{3}~H''\\
	&\;\hspace{2.5cm}+\frac{C_1}{g}\int\frac{dx}{H}+C_\phi
	\Bigg\}\;,\\
w_\psi&=\frac{1}{\sinh(g\sigma)}\Bigg\{
	\frac{\sqrt H H'''}{6\sinh^2(g\sigma)}-\frac{1}{\sqrt H}\left[\frac{C_1}{g}\ln\tanh(g\sigma)+
	\frac{C_H}{6}\left(\frac{1}{2\sinh^2(g\sigma)}+\ln\tanh(g\sigma)\right)\right]\\
	&\;\hspace{2.5cm}+\frac{1}{2\sqrt H}\left[
	(2-C_\phi)x+\frac{2}{3}~H'-\frac{C_1}{g}\int dx\int\frac{dx}{H}+C_\psi
	\right]
	\Bigg\}\;.
\end{aligned}
	\label{eqn:generalw}
\end{equation}
Let us now specialise this expressions for concrete forms of $H(x)$.

\subsubsection{$H(x)$ polynomial}
Set $C_1=0$ in \eqref{eqn:allbhs} and take $H(x)$ to be a  cubic polynomial in $x$ 
\[ H(x)=a_0+a_1x+a_2x^2+a_3x^3 \ . \]
The integrability condition \eqref{hequation} is trivially satisfied since $H''''=0$. In this case, it is easy to see that 
\begin{equation}
\left[(H'''H)'-\frac{(2+H'')^2}{2}\right]'=-2H'''\;,
\end{equation}
and, from \eqref{Ntilde} and \eqref{eqn:allbhs}, we get 
\begin{equation}
N(\sigma,x)=-\frac{g \cosh(g\sigma)\sqrt{H}H'''}{3\sinh^4(g\sigma)}\;.
\end{equation}
This family of solutions includes all known black holes of minimal gauged SUGRA (which form a two parameter family - plus cosmological constant) as well as solutions with non compact horizons, whose spatial sections are homogeneous \textit{Nil} or $SL(2,\mathbb{R})$ manifolds. We can summarise the situation in the following way:
\begin{description}
\item[{\bf Cubic polynomial, $a_3\neq 0$};] Generically the spatial sections of the horizon are non-homogeneous spaces; with appropriate choice of roots it reduces to the supersymmetric $AdS_5$ black hole with two rotation parameters found in 
\cite{Chong:2005hr};
\item[{\bf Quadratic polynomial} $a_3=0$ and $a_2\neq 0$;] The compactness, and even existence, of an horizon depends crucially on the roots. For $a_2<0$, $a_0>0$ the spatial sections of the horizon are the homogeneous $SU(2)$ group manifold and the five dimensional spacetime is the Gutowski-Reall black hole
\cite{Gutowski:2004ez}, which reduces to empty $AdS_5$ when $a_2=-1$, $a_0=1$. For $a_2>0$ the spatial sections of the horizon are the homogeneous $SL(2,\mathbb{R})$ group manifold.
\item[{\bf Linear polynomial} $a_3=a_2=0$;] For all linear polynomials (including constant ones) the spatial sections of the horizon are the \textit{Nil} or Bianchi II group manifold. 
\end{description}
All the solutions with quadratic polynomials were originally found in  \cite{Gutowski:2004ez}, but the ones with non-compact horizons were only studied in the near horizon limit. In our setup we can easily extend them to the whole spacetime.

To justify the statements above note that, for $H''''=0$, 
\begin{equation}
\begin{aligned}
w_\phi&=\frac{1}{\sinh(2g\sigma)}\left\{
	2\sinh^2(g\sigma)-\frac{1}{6\sinh^2(g\sigma)}\left[(H'''H)'-\frac{(2+H'')^2}{2}\right]
	+\frac{1}{3}~H''+C_\phi\right\}\;, \\
w_\psi&=\frac{1}{\sqrt H\sinh(g\sigma)}\left\{
	\frac{H H'''}{6\sinh^2(g\sigma)}+\frac{1}{2}\left[
	(2-C_\phi)x+\frac{2}{3}~H'+C_\psi\right]\right\}\;,
\end{aligned}
\label{eqn:wphiwpsi}
\end{equation}
where $C_\phi$ and $C_\psi$ are integration constants. In the near horizon limit $(\sigma\rightarrow 0)$, therefore,
\begin{equation}
w=-\frac{1}{24g^3\sigma^2}\left[(H'''H)'-\frac{(2+H'')^2}{2}\right](d\phi+xd\psi)+\frac{HH'''}{12g^3\sigma^2}d\psi \ . 
\end{equation}

The metric on the spatial sections of the horizon is 
\[ ds^2=-f^2w^2+f^{-1}ds^2_B|_{\sigma=0} \ .\] 
For a generic {\bf cubic} polynomial, the Ricci scalar of this metric depends on $x$; thus it is a non-homogeneous space. Take the cubic polynomial to be
\begin{equation}
H(x)=\frac{1}{2}(1-x^2)\left[A_1^2+A_2^2+(A_1^2-A_2^2)x\right]\;,
\label{eqn:HPope}
\end{equation}
where $-1\leq x \leq 1$, and $1> A_1^2,A_2^2>0$, which are the two parameters that characterise the solution. From \eqref{eqn:3dsasaki}, one can see that these are the required ranges so that the 3d Sasakian space consists of an $S^1$ fibration over a base $S^2$. Hence, locally, the topology of the horizon is $S^3$. In fact, we get the the near horizon geometry of the general supersymmetric $AdS$ black holes of \cite{Chong:2005hr}. 

The full geometry is obtained as follows. For the 1-form $w$ to be well-defined, the $w_\psi$ component has to vanish on the two poles of the base $S^2$, which implies that $w_\psi$ should be of the form $w_\psi \sim (1-x^2)$. Upon inserting \eqref{eqn:HPope} into  \eqref{eqn:wphiwpsi} and imposing this condition, we can fix $C_\phi$ and $C_\psi$ in terms of the parameters of the K\"ahler base,
\begin{eqnarray}
C_\phi=2-\frac{2}{3}(A_1^2+A_2^2)\;,&\;& C_\psi=\frac{2}{3}(A_1^2-A_2^2)\;. 
\end{eqnarray}
Using these values of $C_\phi$ and $C_\psi$ in \eqref{eqn:wphiwpsi}, we  obtain the final expression for $w_\phi$ and $w_\psi$:
\begin{equation}
\begin{aligned}
w_\phi&=\frac{1}{\sinh(2g\sigma)}\Bigg\{2\sinh^2(g\sigma)+2-(A_1^2+A_2^2)-(A_1^2-A_2^2)x \\
	\;&\hspace{2.5cm}+\frac{1}{3\sinh^2(g\sigma)}
		\Big[1-A_1^2-A_2^2+A_1^4+A_2^4-A_1^2A_2^2-3(A_1^2-A_2^2)x\Big]\Bigg\}\;,\\
w_\psi&=\frac{(1-x^2)(A_1^2-A_2^2)}{2\sqrt H \sinh(g\sigma)}\left\{
	1-\frac{1}{2\sinh^2(g\sigma)}\Big[A_1^2+A_2^2+(A_1^2-A_2^2)x\Big]\right\}\;.
\end{aligned}
\label{eqn:finalw}
\end{equation}

Using the relations given in \eqref{xh} and setting
\begin{eqnarray}
A_1^2=\frac{\Xi_a}{g^2 \alpha^2}\;,&& A_2^2=\frac{\Xi_b}{g^2 \alpha^2}\;,
\end{eqnarray}
with $\Xi_a=1-a^2g^2$, $\Xi_b=1-b^2g^2$ and $\alpha^2=g^{-2}(1+ag+bg)^2$, one can readily check that our solution reproduces the general $AdS$ supersymmetric black hole of \cite{Chong:2005hr}, in the form presented in \cite{Kunduri:2006ek}.\footnote{To see that our solution matches that of \cite{Kunduri:2006ek} one has to set $\beta_2=4r_m^4/3$ in their paper, which is the correct value for the minimal theory. Also note that our solution is rotating in the opposite sense.}

A generic {\bf quadratic} polynomial (i.e. $a_3=0$, $a_2\neq 0$) can be taken in the form $H(x)=a_0+a_2x^2$. The horizon geometry is always homogeneous and can be written 
\begin{equation}
ds^2_H=\frac{1+a_2}{12g^2}\left[\frac{dx^2}{a_0+a_2x^2}+(a_0+a_2x^2)d\psi^2+\frac{1-3a_2}{4}(d\phi+xd\psi)^2\right] \ . \end{equation}
\begin{description}
\item[$\bullet$] If $a_2<0$ (in which case the metric is positive definite for some $x$ domain iff $a_0>0$), define new coordinates $\theta,\tilde{\phi},\tilde{\psi}$ by
\[ x=\sqrt{\frac{a_0}{|a_2|}}\cos\theta \ , \ \ \ \ \psi=\frac{\tilde{\psi}}{\sqrt{a_0|a_2|}} \ , \ \ \ \ \phi=\frac{\tilde{\phi}}{|a_2|} \ , \]
the metric on the spatial section of the horizon can be written as 
\[ ds^2_H=\frac{|a_2|^{-1}-1}{12g^2}\left(d\theta^2+\sin^2\theta d\tilde{\psi}^2+\frac{|a_2|^{-1}+3}{4}(d\tilde{\phi}+\cos\theta d\tilde{\psi})^2\right) \ . \]
This is a standard metric on $SU(2)$ corresponding to a squashed $S^3$. It is the near horizon geometry of the Gutowski-Reall black hole \cite{Gutowski:2004ez} which has only one independent rotation, and is obtained from the generic black hole studied above taking $A_1^2=A_2^2=\frac{1}{4\alpha^2}$ in \eqref{eqn:finalw} and $(\phi,\psi)\to 4\alpha^2(\phi,\psi)$ (to make contact with the form given in \cite{Gutowski:2004ez}). Note that in the limit $a_2\rightarrow 1$, corresponding to taking in the generic black hole  $A_1^2=A_2^2=1$, there is no horizon and we find empty $AdS_5$ (make $g=\frac{\chi}{2\sqrt 3}$ to find the form given in \cite{Gauntlett:2003fk}).
\item[$\bullet$] If $a_2>0$, define new coordinates $y,\tilde{\phi},\tilde{\psi}$ by
\[ {\rm For} \ \ \ \left\{\begin{array}{c} a_0> 0 \\ a_0=0 \\ a_0<0 \end{array} \right. \ , \left\{\begin{array}{c} x=\sqrt{a_0/a_2}\sinh y \\ x=e^y  \\ x=\sqrt{|a_0|/a_2}\cosh y  \end{array} \right. \ ,  \left\{\begin{array}{c} \psi=\tilde{\psi}/\sqrt{a_0a_2} \\\psi=\tilde{\psi}/a_2 \\ \psi=\tilde{\psi}/\sqrt{|a_0|a_2} \end{array} \right. \  ,\ \ \phi=\frac{\tilde{\phi}}{a_2} \ , \]
\end{description}
the metric on the spatial section of the horizon can be written as 
\[ ds^2_H=\frac{a_2^{-1}+1}{12g^2}\left(dy^2+F_1(y) d\tilde{\psi}^2+\frac{a_2^{-1}-3}{4}(d\tilde{\phi}+F_2(y)d\tilde{\psi})^2\right) \ , \]
where
\[ {\rm For} \ \ \ \left\{\begin{array}{c} a_0> 0 \\ a_0=0 \\ a_0<0 \end{array} \right. \ , \ \ \ \ \left\{\begin{array}{l} F_1(y)=\cosh^2y  \ , \  F_2(y)=\sinh y \\ F_1(y)=\exp{2y}  \ , \ F_2(y)=\exp{y} \\  F_1(y)=\sinh^2y  \ , \ F_2(y)=\cosh y  \end{array} \right. \ . \]
These are all metrics on $SL(2,\mathbb{R})$. For instance, if one neglects the conformal factor and uses the triad
\[ {\bf e}^1=dy \ , \ \ \ \ \ {\bf e}^{2}=\sqrt{F_1(y)}d\tilde{\psi} \ , \ \ \ \ \ \ {\bf e}^{3}=\sqrt{\frac{a_2^{-1}-3}{4}}(d\tilde{\phi}+F_2(y)d\tilde{\psi}) \ \ , \]
one checks that for all cases the curvature in an orthonormal frame is
\[ R_{(1)(2)(1)(2)}=-1-\frac{3(a_2^{-1}-3)}{16} \ , \ \ \ \ \ \  R_{(1)(3)(1)(3)}=\frac{1}{4} \ , \ \ \ \ \ \ R_{(2)(3)(2)(3)}=\frac{1}{4} \ . \]

A generic {\bf linear} polynomial (i.e. $a_3=a_2=0$, $a_1\neq 0$) can be taken in the form $H(x)=a_1x$. The horizon geometry is again homogeneous and can be written 
\begin{equation}
ds^2_H=\frac{1}{12g^2}\left[\frac{dx^2}{a_1x}+a_1x d\psi^2+\frac{7}{4}(d\phi+xd\psi)^2\right] \ . \end{equation}
Defining new coordinates
\[
x=\frac{A^2+B^2}{4} \ , \ \ \ \ \psi=\frac{2\arctan{B/A}}{\sqrt{a_1}} \ , \ \ \ \ \phi=\frac{\tilde{\phi}}{2\sqrt{a_1}} \ , \]
the horizon metric is rewritten
\begin{equation}
ds^2_H=\frac{1}{12g^2}\left[dA^2+dB^2+\frac{7}{16a_1}(d\tilde{\phi}+AdB-BdA)^2\right] \ . \end{equation}
the standard homogeneous metric on Nil, or the Heisenberg group. A similarly simple analysis shows that an equivalent metric is still obtained if $H(x)=a_0$. It is now very simple to get the metrics on the whole spacetime for the $SL(2,\mathbb{R})$ case and the $Nil$ using \eqref{eqn:wphiwpsi}.

\subsubsection{$H(x)$ non-polynomial}
Equation \eqref{hequation} has solutions which do not take a polynomial form when $\alpha\neq 0$ or $\beta\neq 0$ in  \eqref{hequationab}. We were not able to find the general solution in this case. Nevertheless we can find some particular solutions, which have, generically, singular horizons.
\begin{description}
\item[For {\bf $\alpha=0, \beta\neq 0$},] a solution takes the form
\[H(x)=kx^{4/3} \ , \ \ \ \ \ k\equiv \left(\frac{3^4\beta}{40}\right)^{1/3} \ . \]
The one-form $w$ can be written, in the near horizon limit, for this case
\[
w=\frac{1}{12g^3\sigma^2}\left[\left(1+\frac{2k}{9x^{2/3}}\right)(d\phi+xd\psi)-\frac{8k^2}{27x^{1/3}}d\psi\right] \ . \]
It is now very simple to compute the metric on the spatial sections of the horizon. Its form is not particularly enlightening; but one verifies that the Ricci scalar depends on $x$ and actually diverges for some positive value of $x=x_0$; for $x>x_0$, the metric is positive definite; the Ricci scalar decreases monotonically from $+\infty$ to a negative value beyond which it increases monotonically approaching zero at $x=+\infty$. The horizon has, therefore, a localised singularity, which makes the solution uninteresting.
\item[For {\bf $\alpha\neq 0$},] a solution takes the form
\[H(x)=k(\alpha x+\beta)^{5/3} \ , \ \ \ \ \ k\equiv \left(\frac{3^4}{40\alpha^4}\right)^{1/3} \ . \]
The one-form $w$ can be written, in the near horizon limit, for this case
\[
w=\frac{1}{12g^3\sigma^2}\left[\left(1+\frac{10k\alpha^2}{9(\alpha x+\beta)^{1/3}}+\frac{10k^2\alpha^4}{3^3(\alpha x+\beta)^{2/3}}\right)(d\phi+xd\psi)-\frac{10k^2\alpha^3}{3^3}(\alpha x+\beta)^{1/3}d\psi\right] \ . \]
Again, it is simple to compute the metric on the spatial sections of the horizon and one finds a similar behaviour to the previous situation. The Ricci scalar depends on $x$ and diverges for some positive value of $x=x_0$; for $x>x_0$, the metric is positive definite; the Ricci scalar decreases monotonically from $+\infty$ to a negative value beyond which it increases monotonically approaching a negative value at $x=+\infty$. Again, the horizon is singular.
\end{description}

\subsection{Solutions with angular dependence}
We now turn to solutions of equations \eqref{eqn:harmonic} which exhibit $\phi$ dependence. We consider explicitly solutions for $\tilde M$ only since solutions for $\tilde N$ will follow in the same way. 

Since $\phi$ parametrises an angular direction, we assume $\tilde M$ to be a periodic function of $\phi$ and hence it should be of the general form
\begin{equation}
\tilde M=\sum_{n\in\mathbb{Z}}\Big[\textstyle a^M_n \cos L_n\phi
	+c^M_n \sin L_n\phi\Big]\; \ , \ \ \ \ \ \ \ L_n\equiv \frac{2\pi n}{\Delta\phi} \ ,  
\end{equation}
where $a^M_n$ and $c^M_n$ may depend on $u,x,\psi$ and $\phi\sim \phi+\Delta\phi$. Plugging this expression into \eqref{eqn:harmonic}  we get the general solution
\begin{eqnarray}
\tilde M&=&\sum_{n\in\mathbb{Z}}\Big\{
	\left[\textstyle A^M_n\cosh L_n u+B^M_n\sinh L_n u\right] 
	\textstyle\cos L_n\phi\nonumber\\
	&\;&\hspace{.9cm}+\left[ \textstyle C^M_n\cosh  L_n u+D^M_n
		\sinh L_n u\right] 
	\textstyle\sin L_n\phi
	\Big\}\;,
\end{eqnarray}
and similarly for $\tilde N$. Here, $A^{M,N}_n$, $B^{M,N}_n$, $C^{M,N}_n$ and $D^{M,N}_n$ are functions of $x$ and $\psi$.

The integrability conditions \eqref{const3}-\eqref{const4} imply
\begin{equation}
\begin{aligned}
A^M_n=-D^N_n\;,&~~~~ B^M_n=-C^N_n\;,&~~~~ C^M_n=B^N_n\;,&~~~~ D^M_n=A^N_n\;,
\end{aligned}
\end{equation}
and hence
\begin{eqnarray}
\tilde N&=&\sum_{n\in\mathbb{Z}}\Big\{
	\left[\textstyle
		D^M_n\cosh L_n u+C^M_n\sinh L_n u\right] 
		\textstyle\cos L_n\phi
	\nonumber\\
	&\;&\hspace{.9cm}-\left[\textstyle B^M_n\cosh L_n u
		+A^M_n\sinh L_n u\right]
		\textstyle\sin L_n\phi
	\Big\}\;.
\end{eqnarray}
From now on we drop the superscript $M$ to avoid cluttering since it should cause no confusion.

The remaining constraints \eqref{const1}-\eqref{const2} yield
\begin{subequations}
\begin{eqnarray}
\partial_x(\sqrt H A^{\pm}_n)\pm \frac{1}{\sqrt H}
	\left[ L_n x~A^{\pm}_n+\partial_{\psi}C^{\pm}\right]&=&0\;,\label{phi1}\\
\partial_x(\sqrt H C^{\pm}_n)\pm \frac{1}{\sqrt H}
	\left[L_n x~C^{\pm}_n-\partial_{\psi}A^{\pm}\right]&=&0\;, \label{phi2}
\end{eqnarray}
\end{subequations}
for $n\neq 0$, together with
\begin{subequations}
\begin{eqnarray}
\frac{\partial_{\psi}A_0}{\sqrt{H}}-\partial_{x}(\sqrt{H}D_0)&=&\frac{g(HH''')'}{3} \ , \label{phi3}\\
 \frac{\partial_{\psi}D_0}{\sqrt{H}}+\partial_{x}(\sqrt{H}A_0)&=&0 \ ,
\label{phi4}
\end{eqnarray}
\end{subequations}
where we have defined $A^{\pm}_n\equiv A_n\pm B_n$ and $C^{\pm}_n\equiv C_n\pm D_n$. These equations are general, but we will focus on $\psi$ independent solutions. We take the functions $A_n^{\pm},C_n^{\pm}$ to have \textit{no} $\psi$ dependence; equations \eqref{phi1}-\eqref{phi2} decouple and have solution
\begin{eqnarray}
A^{\pm}_n=\frac{a^{\pm}_n}{\sqrt H}~\exp\left[\mp L_n\int^x dx'\frac{x'}{H(x')}\right]\;,&\,&
C^{\pm}_n=\frac{c^{\pm}_n}{\sqrt H}~\exp\left[\mp  L_n\int^x dx'\frac{x'}{H(x')}\right]\;, \label{apm}
\end{eqnarray}
where $a^{\pm}_n$ and $c^{\pm}_n$ are constants. Equations \eqref{phi3}-\eqref{phi4}
are solved by
\begin{equation}
D_0(x)=-\frac{g\sqrt{H}H'''}{3}-\frac{C_1}{\sqrt{H}} \; , \ \ \ \ \ \ A_0(x)=0 \ , 
\end{equation}
so that we recover the known solutions setting $n=0$ in the expressions above. Rescaling $a^{\pm}_n$ and $c^{\pm}_n$, we can write down the general solution with $\phi$-dependence:
\begin{eqnarray}
\tilde M&=&\sum_{n\in\mathbb{Z}/\{0\}}\Big[\left(A^+_n e^{L_nu}+A^-_n e^{-L_nu }\right)
	\textstyle\cos\,L_n\phi 
	+\left(C^+_n e^{L_nu }+C^-_n e^{-L_nu }\right)
	\textstyle\sin\,L_n\phi\Big]\;,\\
\tilde N&=&-\frac{g\sqrt{H}H'''}{3}-\frac{C_1}{\sqrt{H}}
	+\sum_{n\in\mathbb{Z}/\{0\}}\Big[\left(C^+_n e^{L_nu}-C^-_n e^{-L_nu}\right)
	\textstyle\cos\,L_n\phi
	\nonumber\\
	&\;&\hspace{5.0cm}-\left(A^+_n e^{L_nu}-A^-_n e^{-L_nu}\right)
	\textstyle\sin\,L_n\phi\Big]\;.\nonumber\\
\end{eqnarray}

The next step is to compute the one-form $w$. Due to our assumption of $\psi$ independence it suffices to consider the following form for $w$
\begin{equation}
w=w_x~{\bf e}^{x}+w_\phi~{\bf e}^\phi+w_\psi~{\bf e}^\psi,
\end{equation}
where the $w_i$ may depend on $\sigma$, $x$ and $\phi$. From \eqref{eqn:dw2} we get,
\begin{eqnarray}
w_x\!&=&\!\frac{\sqrt{e^{-u}-1}}{2g}\bigg\{\sum_{n\in\mathbb{Z}/\{0\}}\frac{1}{L_n}
	\Big[\left(A^+_n e^{L_nu}-A^-_n e^{-L_nu }\right)
	\textstyle\cos\,L_n\phi 	
	\nonumber\\
	&\;&\hspace{4.0cm}+\left(C^+_n e^{L_nu }-C^-_n e^{-L_nu }\right)
	\textstyle\sin\,L_n\phi\Big]
	+k_x(x,\phi)\bigg\}\;,\label{eqn:phiwx}\\
w_\phi\!&=&\!w_\phi^0+\sinh{\left(-u/2\right)}\frac{k_\phi(x,\phi)}{g}\;,
	\nonumber\\
w_\psi\!&=&\!w_\psi^0+
	\frac{\sqrt{e^{-u}-1}}{2g}\bigg\{\sum_{n\in\mathbb{Z}/\{0\}}\frac{1}{L_n}
	\Big[\left(C^+_n e^{L_nu}+C^-_n e^{-L_nu}\right)
	\textstyle\cos\,L_n\phi
	\nonumber\\
	&\;&\hspace{5.0cm}
	-\left(A^+_n e^{L_nu}+A^-_n e^{-L_nu}\right)
	\textstyle\sin\,L_n\phi\Big]
	+k_\psi(x,\phi)\bigg\}\;,\nonumber\\
	\label{eqn:phiwpsi}
\end{eqnarray}
where $w_\phi^0$ and $w_\psi^0$ are given in \eqref{eqn:generalw}. The functions $k_x$, $k_\phi$ and $k_\psi$ obey
\begin{subequations}
\begin{eqnarray}
\partial_\phi k_ x&=&\sqrt H~\partial_x k_\phi\;,\label{eqn:k1}\\ 
\partial_x(\sqrt H~ k_\psi+xk_\phi)&=&0\;, \label{eqn:k2}\\
\partial_\phi(\sqrt H~ k_\psi+xk_\phi)&=&0\;.\label{eqn:k3}
\end{eqnarray}
\end{subequations}
The last two equations require
\begin{equation}
k_\psi=-\frac{x}{\sqrt H}~k_\phi+k\;,
\end{equation}
where $k$ is constant. Therefore, the full solution is given by eqs.\eqref{eqn:phiwx}-\eqref{eqn:phiwpsi}, where  $k_x$ and $k_\phi$ should obey \eqref{eqn:k1}.

\subsubsection{$\phi$ dependent deformations of $AdS_5$}
As a first case study of $\phi$ dependent solutions we take
\[H(x)=1-x^2 \ , \label{hads} \]
together with $C_H=C_1=C_{\psi}=0$ and $C_{\phi}=2/3$ in \eqref{eqn:generalw}. Then $f=1$. We further take $k_{\phi}=k_x=k=0$ as to consider the simplest possible case. Introducing coordinate $x=\cos\theta$ for the base, we get, from \eqref{apm}, 
\[ A^{\pm}_n=a_n^{\pm}(\sin\theta)^{\pm \frac{n}{2}-1} \ , \ \ \ \ C^{\pm}_n=c_n^{\pm}(\sin\theta)^{\pm \frac{n}{2}-1} \ . \]
Thus, putting together \eqref{eqn:generalw} and \eqref{eqn:phiwx}-\eqref{eqn:phiwpsi}, we find
\begin{equation}
w=\frac{\sinh^2g\sigma}{2g}\left(d\phi+\cos\theta\psi\right)+\frac{\sin\theta \Sigma^{\psi}}{4g^2}d\psi-\frac{\Sigma^x}{4g^2}d\theta \ , \label{wdeformedads}\end{equation}
where, noting that the periodicity of $\phi$ is $\Delta\phi=4\pi$ 
\begin{equation}
\begin{array}{l}
\displaystyle{\Sigma^{\psi}=\sum_{n\in\mathbb{Z}/\{0\}}\frac{2}{n\sin\theta}\left[\left(c_n^+\sin^{\frac{n}{2}}\theta\tanh^ng\sigma+c_m^-\sin^{-\frac{n}{2}}\theta\tanh^{-n}g\sigma\right)\cos\frac{n\phi}{2}\right.} \\ \displaystyle{\left. ~~~~~~~~~~~~~~~~~~~~~~~~~~~ -\left(a_n^+\sin^{\frac{n}{2}}\theta\tanh^ng\sigma+a_n^-\sin^{-\frac{n}{2}}\theta\tanh^{-n}g\sigma\right)\sin\frac{n\phi}{2}\right] \ , }\end{array} \end{equation}
\begin{equation}
\begin{array}{l}
\displaystyle{\Sigma^{x}=\sum_{n\in\mathbb{Z}/\{0\}}\frac{2}{n\sin\theta}\left[\left(a_n^+\sin^{\frac{n}{2}}\theta\tanh^ng\sigma-a_n^-\sin^{-\frac{n}{2}}\theta\tanh^{-n}g\sigma\right)\cos\frac{n\phi}{2}\right.} \\ \displaystyle{\left. ~~~~~~~~~~~~~~~~~~~~~~~~~~~ +\left(c_n^+\sin^{\frac{n}{2}}\theta\tanh^ng\sigma-c_n^-\sin^{-\frac{n}{2}}\theta\tanh^{-n}g\sigma\right)\sin\frac{n\phi}{2}\right] \ . }\end{array} \end{equation}
Thus, we find a family of solutions of minimal five dimensional gauged SUGRA of the form
\begin{equation}
ds^2=-[dt+w]^2+ds^2_B \ , \ \ \ \ \ \ \ A=\frac{\sqrt{3}}{8 g^2} \left( \sin\theta \Sigma^{\psi}d\psi-\Sigma^x d\theta \right) \ , \end{equation}where $w$ is given by \eqref{wdeformedads}, and $ds_B^2$ by \eqref{basex} with 
\eqref{hads}. As a simple example take 
\[ c_m^+=\frac{c}{2} \ , \]
for a given positive $m$, and all remaining $c^{\pm}_n=0=a^{\pm}_n$. Then, the full spacetime metric is, explicitly
\begin{equation}
\begin{array}{c} 
\displaystyle{ds^2=-\left[dt+\frac{\sinh^2g\sigma}{2g}\left(d\phi+\cos\theta d\psi\right)+\frac{c\tanh^mg\sigma\sin^{\frac{m}{2}-1}\theta}{4g^2}\left(\sin\theta \cos\frac{m\phi}{2} d\psi-\sin\frac{m\phi}{2} d\theta\right)\right]^2} \\ \displaystyle{~~~~~~~~~~  + d\sigma^2+\frac{\sinh^2g\sigma}{4g^2}\left(d\theta^2+\sin^2\theta d\psi^2+\cosh^2g\sigma(d\phi+\cos\theta d\psi)^2\right)} \ ,\end{array}  \label{deformedads}
\end{equation}
and the gauge field is
\begin{equation}
A=\frac{\sqrt{3}c}{8g^2}\tanh^mg\sigma\sin^{\frac{m}{2}-1}\theta\left(\sin\theta \cos\frac{m\phi}{2} d\psi-\sin\frac{m\phi}{2} d\theta\right) \ . 
\end{equation}
We have verified that these fields obey equations of motion \eqref{eqmotn1d5}.  For $c=0$ they descibe empty $AdS_5$. Generically, this metric is singular for $\sigma=0,+\infty$ and $\theta=0,\pi$. To check if these are physical singularities note that the Ricci scalar takes the form
\[
R=-20g^2+\frac{c^2m^2\left(\sin\theta\sinh^2{g\sigma}\right)^{m-2}}{4\cosh^{2m+2}g\sigma} \ . \]
It does not diverge if $m\ge 2$; otherwise there is a curvature singularity at $\theta=0=\pi$ and at $\sigma=0$. Indeed this seems to be a generic behaviour of the curvature invariants; for instance, the square of the Ricci tensor is 
\[ R_{\mu\nu}R^{\mu \nu}=80g^4-\frac{2g^2c^2m^2\left(\sin\theta\sinh^2{g\sigma}\right)^{m-2}}{\cosh^{2m+2}g\sigma}+\frac{c^4m^4\left(\sin\theta\sinh^2{g\sigma}\right)^{2m-4}}{8\cosh^{4m+4}g\sigma} \ . \]

One typical pathology of this type of solutions is the presence of Closed Timelike Curves. We can show, however, that \eqref{deformedads} is free of Closed Timelike Curves if the parameter $c$ obeys
\begin{equation} c\le 2g \ . \label{noctcs} \end{equation}
To see this, let us analyse the closed directions of the metric $\phi$ and $\psi$. Since $g_{\phi\phi}=\sinh^2g\sigma/4g^2$, $\partial/\partial\phi$ is always spacelike. Since
\[
g_{\psi\psi}=\frac{\sinh^2g\sigma}{4g^2}\left(1-\frac{\chi^2}{\sinh^2g\sigma}\right) \ , \ \ \ \ \ \chi^2\equiv \left(\frac{c}{2g}\right)^2\tanh^{2m}g\sigma\sin^m\theta\cos^2\frac{m\phi}{2} \ , \]
$\partial/\partial \psi$ is always spacelike for the non singular solutions ($m\ge 2$) as long as \eqref{noctcs}. The fact that two given vector fields are spacelike does not guarantee that a linear combination of them is still spacelike, for non-diagonal metrics. This is, in fact, a subtle way in which Closed Timelike Curves may emerge in spacetime. Examples are discussed in \cite{Herdeiro:2000ap,Herdeiro:2002ft,Herdeiro:2003ci,Hubeny:2003sj}. Thus, we must consider the vector field $k=A\partial_{\phi}+B\partial_{\psi}$, which, for appropriately chosen $A,B$ will still have closed orbits. The norm of this vector field can be written
\[
|k|^2=\frac{\sinh^2g\sigma}{4g^2}\left(A^2+B^2\left[1-\frac{\chi^2}{\sinh^2g\sigma}\right]+2AB(\cos\theta-\chi)\right) \ . \]
We could not find any choice of $A,B$ for which this quantity becomes negative. Thus it seems that for \eqref{noctcs} there are no Closed Timelike Curves (or Closed Null Curves) in the spacetime. 

``Static'' coordinates, in the sense that these are the static coordinates of empty $AdS_5$, are obtained introducing a new angular coordinate $\phi'$ and a new radial coordinate $R$ given by
\[ \phi'=\phi-2gt \ , \ \ \ \ \ R=\frac{\sinh g\sigma}{g} \ .  \]
Replacing $\phi,\sigma$ by $\phi',R$ in \eqref{deformedads}, the solutions become time dependent:
\begin{equation}
ds^2=ds^2_{AdS_5}-2\left[(1+g^2R^2)dt+\frac{gR^2}{2}(d\phi'+\cos\theta d\psi)\right]w'-w'^2 \ , \end{equation}
where
\[ ds^2_{AdS_5}=-(1+g^2R^2)dt^2+\frac{dR^2}{1+g^2R^2}+\frac{R^2}{4}\left(d\theta^2+\sin^2\theta d\psi^2+(d\phi'+\cos\theta d\psi)^2\right) \ , \]
\and
\begin{equation}
w'=\frac{c}{4g^2}\left(\frac{g^2R^2}{1+g^2R^2}\right)^m\sin^{\frac{m}{2}-1}\theta\left(\sin\theta \cos\frac{m(\phi'+2gt)}{2} d\psi-\sin\frac{m(\phi'+2gt)}{2} d\theta\right) \ . \end{equation}
The solutions are time dependent in this coordinate chart. Note that $\partial/\partial t$ is an everywhere timelike vector field. Nevertheless, the solution still has, as required by supersymmetry, an everywhere timelike Killing vector field, which is 
$V=\partial_t-2g\partial_{\phi'}$, and in this sense is stationary. In the case of black holes it is with respect to $V$ that there is no ergoregion \cite{Gutowski:2004ez,Hawking:1999dp}. Thinking of $V$ as the generator of time translations corresponds to working in a co-rotating frame, with respect to which these solutions are stationary.  Note also that $V$ becomes null on the conformal boundary of the spacetime. Thus, as for asymptotically $AdS_5$ black holes the boundary is rotating at the speed of light in the co-rotating frame. However, the conformal boundary is different from $AdS_5$, and has metric 
\begin{equation}
\begin{array}{l}
\displaystyle{ds^2=-dt^2+\frac{1}{g^2}\left(d\theta^2+\sin^2\theta d\psi^2+(d\phi'+\cos\theta d\psi)^2\right)}
\\
\displaystyle{-\frac{c}{2g^2}\left[dt+\frac{d\phi'+\cos\theta d\psi}{2g}\right]\sin^{\frac{m}{2}-1}\theta\left(\sin\theta\cos\frac{m(\phi'+2gt)}{2} d\psi-\sin\frac{m(\phi'+2gt)}{2} d\theta\right)} \ . \end{array} \end{equation}
We have checked that this geometry is not conformally flat (Weyl tensor is non-zero). In this sense, the deformations we are considering do not have the same conformal boundary as $AdS_5$; thus they are \textit{not} asymptotically $AdS_5$, and these coordinates should not be dubbed ``static''.

It would be interesting to know the exact amount of supersymmetry these spacetimes preserve (which is at least $1/4$, and cannot be more than $1/2$ \cite{Grover:2006ps}) and compute the conserved quantities associated to the Killing symmetries.\footnote{There is an overlap between the solutions studied in this section and $AdS_5$ deformations found in \cite{Gauntlett:2003fk}. However, therein, these solutions were not analysed, and the method presented here allowed us to deform not only $AdS_5$ but also solutions with different $H(x)$ in a similar fashion. Moreover, the special case with $m=2$ was studied in \cite{Behrndt:2003gc,Gauntlett:2004cm} and it corresponds to a G\"odel type deformation.}

\subsubsection{$\phi$ dependent deformations of the Gutowski-Reall black hole}
We now take,
\be
                      H(x)=A^2(1-x^2) \ ,
\ee
with $0<A^2<1$, corresponding to the Gutowski-Reall black hole. We find
\be
w=\frac{\sinh^2g\sigma}{2g}\left(1+\frac{1-A^2}{2\sinh^2g\sigma}+\frac{(1-A^2)^2}{6\sinh^4g\sigma}\right)\left(d\phi+\cos\theta\psi\right)+\frac{\sin\theta \Sigma^{\psi}}{4g^2}d\psi-\frac{\Sigma^x}{4g^2}d\theta \ , \label{wforGR}
\ee
where, noting that the periodicity of $\phi$ is $\Delta\phi=4\pi/A^2$ 
\begin{equation}
\begin{array}{l}
\displaystyle{\Sigma^{\psi}=\frac{1}{A^2}\sum_{n\in\mathbb{Z}/\{0\}}\frac{2}{n}\left[\left(c_n^+\sin^{\frac{n}{2}}\theta\tanh^{A^2n}g\sigma+c_k^-\sin^{-\frac{n}{2}}\theta\tanh^{-A^2n}g\sigma\right)\frac{\cos\frac{A^2n\phi}{2}}{\sin\theta}\right.} \\ \displaystyle{\left. ~~~~~~~~~~~~~~~~~~~~~ -\left(a_n^+\sin^{\frac{n}{2}}\theta\tanh^{A^2n}g\sigma+a_n^-\sin^{-\frac{n}{2}}\theta\tanh^{-A^2n}g\sigma\right)\frac{\sin\frac{A^2n\phi}{2}}{\sin\theta}\right] \ , }\end{array} \end{equation}
\begin{equation}
\begin{array}{l}
\displaystyle{\Sigma^{x}=\frac{1}{A^2}\sum_{n\in\mathbb{Z}/\{0\}}\frac{2}{n}\left[\left(a_n^+\sin^{\frac{n}{2}}\theta\tanh^{A^2n}g\sigma-a_n^-\sin^{-\frac{n}{2}}\theta\tanh^{-A^2n}g\sigma\right)\frac{\cos\frac{A^2n\phi}{2}}{\sin\theta}\right.} \\ \displaystyle{\left. ~~~~~~~~~~~~~~~~~~~~~ +\left(c_n^+\sin^{\frac{n}{2}}\theta\tanh^{A^2n}g\sigma-c_n^-\sin^{-\frac{n}{2}}\theta\tanh^{-A^2n}g\sigma\right)\frac{\sin\frac{A^2n\phi}{2}}{\sin\theta}\right] \ . }\end{array} \end{equation}
Taking 
\[ c_m^+=\frac{cA^2}{2} \ , \]
for a given positive $m$, and all remaining $c^{\pm}_n=0=a^{\pm}_n$, the full spacetime metric is of the form \eqref{timelikesolution} with 
\be\begin{split}
                        w=&\frac{\sinh^2g\sigma}{2g}\left(1+\frac{1-A^2}{2\sinh^2g\sigma}+\frac{(1-A^2)^2}{6\sinh^4g\sigma}\right)\left(d\phi+\cos\theta d\psi\right)\\[6pt]
                          &+\frac{c\tanh^{A^2m}g\sigma\sin^{\frac{m}{2}-1}\theta}{4g^2}\left(\sin\theta \cos\frac{A^2m\phi}{2} d\psi-\sin\frac{A^2m\phi}{2} d\theta\right) \ ,
\end{split}\ee
\be
                         f^{-1}=1+\frac{1-A^2}{3\sinh^2g\sigma} \ ,
\ee
and
\be
                      ds^2_B(\mcK)=d\sigma^2+\frac{\sinh^2g\sigma}{4g^2}\left(\frac{d\theta^2}{A^2}+A^2\sin^2\theta d\psi^2+\cosh^2g\sigma(d\phi+\cos\theta d\psi)^2\right)\ ,
\ee
while the gauge field is
\be\begin{split}
                      A=&\frac{\sqrt{3}}{2}f\,dt+\frac{\sqrt{3}}{2}\left(1+2\frac{1-A^2}{3\sinh^2g\sigma}+\frac{(1-A^2)^2}{2\sinh^4g\sigma}+\frac{(1-A^3)^2}{18\sinh^6g\sigma}\right)\frac{\sinh^2g\sigma}{2g}(d\phi+\cos\theta\,d\psi)\\[10pt]
                      &+\frac{\sqrt{3}c}{8g^2}\left(1+\frac{1-A^2}{3\sinh^2g\sigma}\right)\tanh^{A^2m}g\sigma\sin^{\frac{m}{2}-1}\theta\left(\sin\theta \cos\frac{A^2m\phi}{2} d\psi-\sin\frac{A^2m\phi}{2} d\theta\right) \ .
\end{split}\ee
These deformations of the Gutowski-Reall black hole preserve $1/4$ of the supersymmetry.\footnote{As for the empty $AdS_5$ case the G\"odel type deformation ($m=2$) was previously studied in \cite{Behrndt:2004pn}.} For any positive $m$ the Ricci scalar (and presumably the other curvature invariants) does not diverge at the horizon. In fact, the deformation terms are subleading corrections, at the horizon, both in the metric and in the curvature invariants. The analysis of the asymptotic structure is analogous to the previous section. 

A similar analysis for all other solutions of \eqref{hequation} could now be done, including the most general $AdS_5$ black holes \eqref{eqn:HPope}.

\section{Conclusions and Discussion}
The main goal of this paper was to investigate more general supersymmetric black hole solutions in $AdS_5$ than the ones known hitherto. With this purpose, we constructed a family of K\"ahler bases using some assumptions that are compatible with the existence of an event horizon. These assumptions are restrictive and, from the outset, our analysis did not cover possible $AdS_5$ black rings. Nevertheless the analysis, and the family of K\"ahler bases that emerged from it, still proved fruitful. 

Firstly, the function that characterises the family of bases, obeys a remarkably simple 6th order differential equation. A family of solutions, albeit not the most general one, is a cubic polynomial, which turns out to describe all known $AdS_5$ supersymmetric black holes. The remaining solutions of the sixth order equation that we found are spaces with non-compact horizons or spaces with singular horizons. This suggests that indeed, the family of solutions found so far describes the most general black holes with spherical topology and two axisymmetries in $AdS_5$. More general black holes, with fewer isometries are not excluded and seem compatible with the results in \cite{Hollands:2006rj}. However, following a similar analysis to the one herein for less symmetric base spaces does not seem as tractable a problem and presents a challenge. In the appendix we make some preliminary analysis in that direction.

Secondly, we found an infinite set of supersymmetric deformations of both $AdS_5$ space and the black holes living on it. These deformations vanish at the horizon and they change the asymptotic structure of the spacetime.  Interestingly, these deformations, are time dependent, when one writes $AdS_5$ in static coordinates, but there is still an everywhere timelike Killing vector field. Also, they provide an example of supersymmetric $AdS_5$ solutions, where the five dimensional spacetime has less spatial isometries than the base space. Of course it would be interesting to understand better these solutions and if they have some CFT correspondence. Let us note that a similar set of deformations, albeit more restrictive, was studied for the ungauged theory in \cite{Ortin:2004af}, where it was shown that any asymptotically flat solution could be embedded in a G\"odel type universe. This is the vanishing cosmological constant limit of the $m=2$ deformation presented herein. All remaining regular deformations vanish in that limit. 

There are several directions in which one could extend the work herein. One of the outstanding questions in this field, is the possible existence of $AdS_5$ black rings. In order to consider black rings, an analysis similar to the one in section \ref{conical} could, perhaps, be followed,  replacing \eqref{nonsing} by something of the type
\be
ds_B^2(\mcK)\simeq ds^2_2(\rho,\theta) + f_{11}(\theta)\rho^2\,(d\phi^{1})^2+ f_{22}(\theta)R^2\,(d\phi^{2})^2+ f_{12}(\theta)\rho R\,d\phi^{1}d\phi^{2}\ ,     \label{nonsingring}
\ee 
where $R$ is constant. But losing the property that the base of the solution is a K\"ahler cone in the vicinity of a possible horizon makes the analysis much harder. One other possible generalisation of our study is to seek other $G(\rho)$ that approach $\rho^2$ in the vicinity of a horizon and become asymptotically Bergmann. We have not been able to find other interesting examples.

Finally, let us mention that the issue of the existence of multi-black hole solutions in $AdS_5$ remains open. Clearly, a strategy to look for such solutions is to consider a K\"ahler base that reduces to a conical K\"ahler manifold at a set of points. However, one expects fewer isometries in such case, making the problem more difficult, since the base is not toric. Still, until a physical argument is given excluding the possibility of having multi-black hole solutions in $AdS_5$ this seems an interesting open question.

\section*{Acknowledgements}

We are very grateful to Roberto Emparan, Jan Gutowski and Harvey Reall for discussions and suggestions on drafts of this paper. C.H. would also like to thank J. Lucietti and H.Kunduri for an interesting conversation. P.F would like to thank Centro de F\'\i sica do Porto for hospitality, during the early stages of this work. C.H. and F.P.C are supported by FCT through the grants SFRH/BPD/5544/2001 and SFRH/BPD/20667/2004. P.F. is supported by a FI fellowship of the DURSI (Generalitat de Catalunya) and by the grants DURSI 2005 SGR 00082, CICYT FPA 2004-04582-C02-02 and EC FP6 program MRTNCT-2004-005104. This work was also supported by Funda\c c\~ao Calouste Gulbenkian through \textit{Programa de Est\'\i mulo \`a Investiga\c c\~ao}, by the FCT grants POCTI/FNU/38004/2001 and POCTI/FNU/50161/2003. Centro de F\'\i sica do Porto is partially funded by FCT through POCTI programme.

\appendix

\section{Non-toric K\"ahler cones}\label{nontoric}
Equation \eqref{hequation} was obtained for toric K\"ahler cones, admitting a $U(1)^2$ action. Some such cones provide the base (near the horizon) for black holes admitting a $U(1)^2$ spatial isometry. If one would like to search for black objects with a smaller spatial isometry group one could consider cones with less symmetry. So, we consider now the possibility that the 2d base of the Sasakian space is not toric. From the discussion in section \ref{fibration} we can see that the Sasakian space still has one isometry, as $\eta=d\phi+\xi$ where $\xi$ is parallel to the 2d base. Thus, the four dimensional cone will only admit a one dimensional isometry group. Let us derive, in this more general case, what is the constraint analogous to \eqref{hequation}.

To study the near-horizon geometry of 5d solutions having non-toric K\"ahler
cones as their bases, let us define these bases by the following set of 1-forms:
\be
                  {\bf e}^\rho=d\rho \ ,\quad {\bf e}^\eta=\rho\eta \ ,\quad {\bf e}^w=\rho\sqrt{H}dw \
,
\ee
where we introduced the function
\be
                  H(w,{\bar w})\equiv {\hat \mcK}_{w{\bar w}} \ .
\ee
In this basis the K\"ahler form of the cone and $G^+$ read
\be
                  J=-({\bf e}^\rho\wedge {\bf e}^\eta+i{\bf e}^w\wedge {\bf e}^{\bar w}) \ ,
\ee
and
\be
                  G^+=-3gf^{-1}(w,{\bar w})({\bf e}^\rho\wedge {\bf e}^\eta-i{\bf e}^w\wedge
{\bf e}^{\bar w}) \ .
\ee
The most general ansatz for $G^-$ then reads
\be
                  G^-=-LJ-f(T\,{\bf e}^\rho\wedge
{\bf e}^w+\textup{c.c.})-f(iT\,{\bf e}^\eta\wedge {\bf e}^w+\textup{c.c.}) \ .
\ee
We find easily that \eqref{eq:Gsquared} is still valid. We obtain therefore
\be
                  f^{-1} L(\rho, w,{\bar w})=\frac{1}{4g}\nabla^2
f^{-1}=\frac{1}{g\rho^4 2H}\partial_w\partial_{\bar w} c(w,{\bar w}) \ .
\ee
The complex function $T$ is determined by imposing the integrability condition
\eqref{closed}. For the sake of simplicity, we solve these constraints
assuming that $T=T(\rho,w,{\bar w})$. We obtain in this way the following set
of equations:
\be
  i\partial_\rho \rho^2 T - \partial_\phi
T-\frac{1}{\rho^3\sqrt{H}}\partial_w\left(c\,l-3gc^2\right)=0 \ ,
\ee
\be\label{eq:secondconstraint}
                    \partial_w(\sqrt{H}{\bar T})-\partial_{\bar
w}(\sqrt{H}{T})-i4H\frac{c\,l}{\rho^4}=0 \ ,
\ee
\be
                     \partial_w(\sqrt{H}{\bar T})+\partial_{\bar
w}(\sqrt{H}{T})=0 \ .
\ee
The general solution is
\be
                    T=\frac{i}{2\rho^4\sqrt{H}}\partial_w\left(c\,l-3gc^2\right)+\frac{1}{\rho^2\sqrt{H}}T_0(w)
\ ,
\ee
where $T_0(w)$ is holomorphic. Plugging this back in \eqref{eq:secondconstraint}
we obtain a constraint on the geometry of the base, namely
\be
                     \partial_w\partial_{\bar w}\left( c\,l-3gc^2+\frac{2}{g}c
\right)=0 \ ,\label{wwequation}
\ee
where
\be
                      c\,l=\frac{1}{2gH}\partial_w\partial_{\bar w}c \ ,
\ee
and
\be
                      c=\frac{1}{3g^2}+\frac{1}{12g^2H(w,{\bar w})} \partial_w
\partial_{\bar w} \ln{H(w,{\bar w})} \ .
\ee
Equation \eqref{wwequation} is a generalisation of \eqref{hequation}, to which it reduces for
toric K\"ahler cones. It can be easily reduced to a 4th order non-linear
partial differential equation which reads
\be
                      c\,l-3gc^2+\frac{2}{g}c =h(w)+{\bar h}({\bar w}) \ .
\ee

Another way of writing the 6th order equation is
\be
                      \nabla_2^2 \left(\nabla_2^2 R_2+\half(R_2)^2\right)=0 \ ,
\ee
where $\nabla_2^2$ and $R_2$ are the Laplacian operator and the scalar curvature
on the 2d base space; the latter can be expressed as
\be 
R_2=-\frac{2}{H}\partial_{w}\partial_{\bar{w}}\ln H \ . \ee 
Requiring this space to be compact implies that
\be
                      \nabla_2^2 R_2+\half(R_2)^2=const \ ,
\label{constr_curvature}\ee
for in this case any harmonic function is constant. The trivial solutions
$R_2=const > 0$ lead to empty $AdS_5$ and the black hole of Gutowski and Reall.
For general toric cones this equation reduces to $H^2H''''=const$. Note that
$R_2=0$ is also a solution, showing that \eqref{constr_curvature} does not
necessarily lead to compact spaces.

\end{document}